\documentclass[12pt]{article}
\usepackage{fullpage}
\title{BRST APPROACH TO HAMILTONIAN SYSTEMS}
\author{A.K.Aringazin$^1$, V.V.Arkhipov$^2$, and A.S.Kudusov$^3$ \\
                Department of Theoretical Physics\\
                  Karaganda State University\\
                Karaganda 470074 Kazakstan }
\date{Preprint KSU-DTP-10/96}
\begin{document}
\begin{titlepage}
\maketitle
\abstract
{BRST formulation of cohomological Hamiltonian mechanics is
presented.
 In the path integral approach, we use the BRST gauge fixing
procedure for the partition function with trivial underlying
Lagrangian to fix symplectic diffeomorphism invariance.
 Resulting Lagrangian is BRST and anti-BRST exact and the
Liouvillian of classical mechanics is reproduced in the
ghost-free sector.
 The theory can be thought of as a topological phase of Hamiltonian
mechanics and is considered as one-dimensional cohomological
field theory with the target space a symplectic manifold.
 Twisted (anti-)BRST symmetry is related to global $N=2$
supersymmetry, which is identified with an exterior algebra.
 Landau-Ginzburg formulation of the associated $d=1$, $N=2$ model
is presented and Slavnov identity is analyzed.
 We study deformations and perturbations of the theory.
 Physical states of the theory and correlation functions of
the BRST invariant observables are studied.
 This approach provides a powerful tool to investigate
the properties of Hamiltonian systems.

PACS number(s): 02.40.+m, 03.40.-t,03.65.Db, 11.10.Ef, 11.30.Pb.}

\vfill

\noindent
\hbox to 3cm{\hrulefill}\\
$^1${ascar\mbox{@}ibr.kargu.krg.kz}\\
$^2${arkhipov\mbox{@}ibr.kargu.krg.kz}\\
$^3${kudusov\mbox{@}ibr.kargu.krg.kz}

\end{titlepage}

\section{INTRODUCTION}\label{introduction}

 Recently, path integral approach to classical mechanics
has been developed by Gozzi, Reuter and Thacker in a series
of papers\cite{GR-89}-\cite{G-93}.
 They used a delta function constraint
on phase space variables to satisfy Hamilton's equation and a sort
of Faddeev-Popov representation.
 This constraint has been exponentiated with the help of Lagrange
multiplier, ghost and anti-ghost fields so that the resulting field
theoretic Lagrangian appears to be BRST and anti-BRST invariant.

 This is quite analogous to the usual path integral
formulation of quantum gauge field theories, in which BRST
symmetry of the gauge fixed Lagrangian has been originally
found.
 Due to the standard Faddeev-Popov procedure, one starts with
a classical Lagrangian, which is invariant under the action of
a gauge group, and the gauge fixing yields additional
(gauge fixing and ghost dependent) terms in the Lagrangian.
 The BRST symmetry of the resulting gauge fixed Lagrangian
is well known to be a fundamental property providing, particularly,
renormalizability of the theory.

 Wellknown alternative method to quantize gauge field theories is
just based on the BRST symmetry. Instead of implementing the
gauge fixing constraint, one simply insists on the BRST invariance
from the beginning, by constructing nilpotent BRST operator and
BRST exact Lagrangian.
 The BRST quantization scheme provides a simple geometrical basis
for heuristic Faddeev-Popov method and is known as a powerful
tool to deal with not only gauge field theories but also with much
more complicated field theories.
 An important point is to identify the symmetries to be fixed.

 Trivial Lagrangians are known to be of much importance in the
cohomological quantum field
theories\cite{Witten-88}-\cite{Dubrovin-92}.
 As it was realized, these theories can be derived by an
appropriate BRST gauge fixing of a theory in which the underlying
Lagrangian is zero.
 An extensive literature exists on the topological field theories.
 Various topological quantum field theories, such as
topological Yang-Mills theories,
two-dimensional gravity\cite{Witten-90-NP}-\cite{Perry-92/33},
four-dimensional conformal gravity\cite{Perry-92/42},
non-linear sigma model\cite{Witten-88a},
Lan\-dau-Ginz\-burg models\cite{Vafa-91},
two-dimensional BF models\cite{Guadagnini-91},
WZNW models\cite{Hwang-93},
$W$-strings\cite{Bergshoeff-92}
are investigated within the BRST quantization scheme; see
Ref.\cite{Birmingham-91} for a review.
 We feel it is worthwhile to broaden this effort and, in this paper,
use the BRST procedure to develop a model describing classical
dynamical systems.

 The model described by the partition function (\ref{Z}) is by
construction one-dimensional cohomological field theory, in the
sense that the resulting Lagrangian is BRST exact, with trivial
underlying Lagrangian.
 The theory (\ref{Z}) can be thought of as a topological phase of
Hamiltonian mechanics.
 The resulting Lagrangian is in effect the same one obtained by
Gozzi, Reuter and Thacker plus additional $\alpha$ dependent terms,
which we drop in the subsequent consideration.
 We should emphasize here that, stating that the theory is,
as such, a sort of topological one, they proved\cite{GR-90a} that
the partition function is proportional to Euler characteristic of
the phase space and studied $2n$-ghost ground state sector of
the associated supersymmetric model.
 However, more elaborated analysis is needed.
 In the present paper, we use the tools of topological field
theory to fill this gap.

 The most close examples of cohomological field theory to the one
considered in this paper are topological nonlinear sigma
model\cite{Witten-88a}, in which a basic field is the map from
Riemannian surface to a fixed Kahler manifold as the target space,
and topological Landau-Ginzburg models\cite{Vafa-91}.
 Also, one-dimensional ($d=1$) sigma model with the target space
a compact Kahler manifold having nontrivial homotopy group
$\pi_{1}$ has been considered recently by Cecotti and
Vafa\cite{Cecotti-92}, in connection with the Ray-Singer analytic
torsion.

 One of the themes underlying this paper is the notion that
studying topological field theoretic models has regularly
proved useful in developing our understanding of field theories
and physical phenomena more generally.
 For example, studying cohomological content of $d=2$, $N=2$
supersymmetric model has enhanced understanding of
two-dimensional Ising model\cite{Cecotti-92}.
 Despite the fact that various topological field theories have been
thoroughly studied, we think it is useful to look closely at
the specific model, which has its own significance in the context
of continued studies of classical dynamical systems.
So, we apply various tools and analyze the model in many
directions.

 Since we need to fix only one symmetry, the problem of giving
the BRST formulation itself of the cohomological
classical mechanics which we develop in this paper is drastically
simpler than that of aforementioned topological field theories,
which are characterized by rich field content and a set of
symmetries.
 Nevertheless, we give a systematic representation in order to
provide a self-consistent and precise description.
 Also, the theory is one-dimensional that makes general structure
of the theory less complicated than that of the two- or higher
dimensional topological field theories; for example, there are
no conventional "instanton corrections" and the notion of spin
is irrelevant.
 Furthermore, it should be stressed that the resulting theory is
essentially a {\sl classical} one despite the fact that we are
using the BRST gauge fixing scheme, which is usually exploited
to construct quantum field theories.
 As the result, sharing many properties with the usual topological
quantum field theories, it differs from those by absence
of quantum ($\hbar$) corrections.
 However, we should emphasize that there arises a major
distinction from the conventional topological field theories
studied in the literature since symplectic structure of the target
space rules out quadratic terms from the Lagrangian, leading to a
first-order character of the system.
 Such specific theories, describing Hamiltonian systems, are worth
to be studied exclusively.

 In view of the above, cohomological classical mechanics represents
perhaps the simplest example of topological field theory.
 So, after constructing the BRST invariant Lagrangian the focus
of our paper is to study the implications and novel aspects
arising from the BRST approach and associated supersymmetry.
 The work in this paper enables us to use supersymmetric field
theory as a way of deeper understanding of Hamiltonian systems.
 In general, this approach provides a powerful tool to investigate
the fundamental properties and characteristics of Hamiltonian
systems such as ergodicity, Gibbs distribution,
Kubo-Martin-Schwinger condition\cite{GR-90}, integrability,
and Lyapunov exponents\cite{GR-92}.
 Particularly, we think it is illuminating and instructive to map
out some identifications one can draw between the topological field
theories and Hamiltonian systems.
 This is, in part, to establish dictionary between the old and
modern techniques used in studying classical dynamical systems.

 Another motivation of our study is that we consider the BRST
formulation of cohomological classical mechanics as providing
a basis to give BRST formulation of (cohomological)
quantum mechanics and, from then on, apply topological
field theory methods to study quantum mechanical systems.
 The key to making the connection between them lies
in treating quantum mechanics as a smooth $\hbar$-deformation
of the Hamiltonian one, within the phase space
(Weyl-Wigner-Moyal) formulation of quantum mechanics\cite{GR-92d}.
 We hope, in this way, that one might investigate quantum
ergodicity and quantum chaos characteristics which are now of
striking interest.

 In addition, there arises a tempting possibility to give a
classification of possible topologies of constant energy
submanifolds of the phase space for the case of reduced Hamiltonian
systems.
 Of course, this idea is reminiscent of the one of using BRST
symmetry and supersymmetry to obtain various topological results.
 Indeed, we already know that the instantons and Witten index
serve as the tool to obtain such quantities as the Donaldson
invariants, Lefschetz number, and Euler characteristic.
 The idea to combine the tools of topological field theories and
classical Morse theory might be productive here as well.

 As to examples of field theoretic approach to Hamiltonian
systems we notice that the path integral approach to
Euler dynamics of ideal incompressible fluid viewed as Hamiltonian
system has been developed recently by Migdal\cite{Migdal-93},
to study turbulence phenomenon in terms of the path integral over
the phase space configurations of the vortex cells.
 Hamiltonian dynamics has been used to find an invariant
probability distribution which satisfies the Liouville equation,
with topological terms in the effective energy being of much
importance.

 Also, more recently Niemi and Palo\cite{Niemi-94} considered
classical dynamical systems using $d=2$, $N=2$ supersymmetric
nonlinear sigma models. They followed studies on the Arnold
conjecture on the number of $T$-periodic trajectories\cite{Hofer-94}
by Floer\cite{Floer-86}, who proved the conjecture for the
symplectic manifolds subject to the condition that the integral
of symplectic two-form over every two-dimensional sphere is zero.
 Particularly, they used a generalization of Mathai-Quillen
formalism, previously applied in the investigation of Witten's
topological sigma model, and studied functional Hamiltonian flow in
the space of periodic solutions of Hamilton's equation by breaking
the (1,1) supersymmetry with Hamiltonian flow down to a chiral
(1,0) supersymmetry to describe properties of the action of
the model in terms of (infinite dimensional) Morse theory.

 The outline of the paper is as follows.

 In Sec.~\ref{brst}, we start with a target space interpretation
of Hamiltonian mechanics and explore the BRST gauge fixing scheme
to fix diffeomorphism invariance of the trivial underlying
Lagrangian (Sec.~\ref{brst1}).
 The gauge fixing condition is Hamilton's equation plus some
additional $\alpha$ dependent term.
 When both the BRST and anti-BRST symmetries are incorporated
there appears no room for the $\alpha$ dependent terms
in the Lagrangian, which exhibits $Z_2$ symmetry.
 The Liouvillian of ordinary classical mechanics is reproduced
by the associated Hamilton function, in the ghost-free sector.
 The model reveals symplectic structure represented by using of the
cotangent superbundle over phase space naturally supplied by the
field content.
 Then we analyze Slavnov identity to demonstrate that the model is
perturbatively trivial and BRST anomaly free (Sec.~\ref{Slvnv}).

 With this set up, in Sec.~\ref{twist} we study in some detail
the associated supersymmetric model and cohomology.
 Namely, we use topological twist of the BRST and anti-BRST
operators to obtain global $N=2$ supersymmetry\cite{GR-90}
(Sec.~\ref{Ssym}),
and relate the supersymmetry to an exterior algebra
(Sec.~\ref{Chmlg}). In so doing, we are able to identify physical
states of the theory.
 The link between the BRST and anti-BRST symmetries is the
supersymmetric ground state sector, {\sl i.e.} the Ramond sector,
of the associated $d=1$, $N=2$ model.
 Due to the underlying supersymmetry of the Hamilton function
(\ref{H}), only the Ramond states are of relevance which are
found to be in correspondence with cohomology classes of the
target manifold. The states can be in general treated as
(cohomology classes of) form valued classical probability
distributions on the phase space $M^{2n}$.
 In the ghost-free sector, they correspond to the probability
distribution related\cite{GR-90} to conventional ergodic Hamiltonian
systems.
 Physically relevant (normalizable) solutions of the
supersymmetry equations are given specifically by the Gibbs state
form coming from the $2n$-ghost sector.
 We stress that the supersymmetry appears to be a strong
constraint on the physics.
 Criterion for regular/nonregular motion regimes
in Hamiltonian systems is related to the Witten index
known as a measure for supersymmetry breaking.
 Partition function evaluates Euler characteristic of the target
space and the Witten index is equal to Euler characteristic, too
(not surprising result, certainly, obtained earlier in the context
of topological $d=2$, $N=2$ sigma models).
 Also, we find that Poincare integral invariants can be naturally
identified as homotopically trivial BRST invariant observables.
 Existence of homotopically nontrivial Poincare invariants
is a consequence of the field theoretic approach.

 We discuss briefly on the connection of the model to Morse theory
(Sec.~\ref{Mrs}) observing that Hamiltonian may serve as a
Morse function and then proceed to obtain Landau-Ginzburg
formulation of the $d=1$, $N=2$ model using a superspace
technique (Sec.~\ref{LG}). One of the results is that the model
admits Landau-Ginzburg description so that its properties can be
largely understood in terms of superpotential.
 The lowest component of the superpotential has been identified
as Hamiltonian. The action appears to be in the form of a D-term.
 The ring of chiral operators consists of polynomials modulo
the relation characterizing critical points of the Hamiltonian
flow.

 We show that the time reparametrization invariance of the model
requires the fundamental homotopy group $\pi_1(M^{2n})$ to be
nontrivial.

 Valuable information comes from studying possible deformations
and perturbations of the action. We analyze deformations of the
superpotential and symplectic tensor (Sec.~\ref{Dfrm}).
 What is most interesting is that the supersymmetry preservation
condition for the deformation of superpotential (Hamiltonian)
by analytic function is explicitly related to integrals of motion.
 This is another step toward revealing connection between the
supersymmetry and integrability properties of the system.
We also study a deformation of the coordinate dependent
symplectic tensor for which case slight modifications of the
BRST structure have been accounted. A remarkable result is that
to preserve the supersymmetry Schouten bracket between the
deformation tensor and symplectic tensor must be zero.
Also, we use a generalized Mathai-Quillen formalism to construct
the action in terms of an equivariant exterior derivative in the
space of fields (Sec.~\ref{Eqvr}) to get a more clear geometrical
meaning of the model and to provide a possible set up for studying
supersymmetry breaking. This result may serve as a foundation for
further work.

 In Sec.~\ref{correlation}, we study BRST invariant observables
and its correlation functions.
 The BRST invariant observables of interest are closed $p$-forms
on symplectic manifold and correspond to the de Rham cohomology
classes (Sec.~\ref{brstobs}).
 Elaborating connection between the BRST symmetry and
supersymmetry, we identify the BRST invariant observables
with chiral operators of the $d=1$, $N=2$ model.
 The anti-BRST observables can be treated in the same manner.

 Also, there arise naturally homotopy classes of classical
periodic orbits (Sec.~\ref{hmtp}) so that coefficients of the
$p$-forms take values in the linear bundles of appropriate
representations of $\pi_1(M^{2n})$.
 This leads to consideration of the loop space consisting of
mappings $S^1 \to M^{2n}$ which is a natural object in the
topological framework. We remark that periodic orbits are
in many ways the key to the classical dynamics.
Using the above mapping and an integer valued closed two-form,
we construct a term which can be added to the original action
and provides a possible mechanism for symmetry breaking.

 The correlation functions (Sec.~\ref{corfun})
are found to be related to the
intersection number, with the two-point correlation function,
in the homotopically trivial sector, representing the standard
intersection form in cohomology identified as the topological
metric.
 As it is in the topological field theories, there are no
"local" degrees of freedom in the theory under consideration
that means that the correlation functions are not time dependent.

 A certain kind of correlation functions intertwines the BRST and
anti-BRST sectors and are known to be related to the Lyapunov
exponents, positive values of which are strong indication of chaos
in Hamiltonian system.
 The $p$-form sectors, $Z_{p}$, $p=0,\dots,2n$, of the partition
function (\ref{Z}) for the periodic orbits are evaluated via
realizing $M^{2n}$ as a covering space and using monodromy.

 In Sec.~\ref{discussion}, we end the paper with some comments
about what questions one might address next.

\section{BRST FORMULATION}\label{brst}

 A Hamiltonian dynamical system can be described geometrically by a
phase space manifold $M^{2n}$ equipped by a symplectic form
$\omega$ and a Hamiltonian $H$\cite{Abraham-Arnold}.
 The evolution of the system in time $t$ is given by
a particular set of trajectories on $M^{2n}$, parametrized by $t$,
such that Hamilton's equation holds.

 On the other hand, Hamiltonian dynamical system can be described as
one-dimensional field theory, in which dynamical variable is the map,
$a^{i}(t): M^1 \rightarrow M^{2n}$, from one-dimensional space $M^1$,
$t \in M^1$, to $2n$-dimensional symplectic manifold $M^{2n}$.

 The commuting fields $a^{i}= (p_1,\dots ,p_n , x^1,\dots ,x^n)$
are local coordinates on the target space, the phase space $M^{2n}$,
endowed with a nondegenerate closed two-form $\omega$;
$\omega^n \not = 0$, $d\omega = 0$.
In terms of the local coordinates,
$\omega = \frac{1}{2}\omega_{ij}da^i\wedge da^j$;
$\omega_{ij}\omega^{jk} = \delta_i^k$.
 We take $\omega_{ij}$ to be constant symplectic matrix,
{\sl i.e.} use canonical (Darboux) coordinates.

\subsection{BRST APPROACH TO HAMILTONIAN SYSTEMS}\label{brst1}

 Our starting point is the partition function
\begin{equation}
                                                      \label{Z0}
Z = \int D[a]\ \exp iI_{0},
\end{equation}
where $I_{0} = \int dt {\cal L}_{0}$ and the Lagrangian
is trivial, ${\cal L}_{0}= 0$.
 This Lagrangian has symmetries more than the usual diffeomorphism
invariance.

 The BRST gauge fixing
scheme\cite{Birmingham-91,Baulieu-88}
assumes fixing of some symmetry of the underlying action by
introducing appropriate ghost and anti-ghost fields.
 The symmetry of the action (\ref{Z0}) we are interested in is
symplectic diffeomorphism invariance, which leaves the symplectic
tensor $\omega_{ij}$ form invariant,
\begin{equation}
                                                          \label{a}
\delta a^{i} = \ell_{h}a^{i}.
\end{equation}
 Here, $\ell_{h} = h^{i}\partial_{i}$ is a Lie-derivative along
the Hamiltonian vector field,
$h^{i} = \omega^{ij}\partial_{j}H(a)$\cite{Abraham-Arnold}.
 The transformations (\ref{a}) are canonical ones, which preserve
the usual Poisson brackets, $\{,\}_\omega$, defined by the
symplectic tensor. Note, however, that the symmetry under (\ref{a})
can be viewed more generally as an invariance under any
diffeomorphism, with $h$ treated as a vector field.

 By introducing the ghost field $c^{i}(t)$ and the anti-ghost
field $\bar c_{i}(t)$, we write the BRST version of the
diffeomorphism (\ref{a}),
\begin{equation}
                                                         \label{s}
sa^{i}= c^{i},\quad
sc^{i}= 0, \quad
s \bar c_{i}= iq_{i},\quad
sq_{i}= 0,
\end{equation}
where the BRST operator $s$ is nilpotent, $s^2=0$, and $q_{i}$ is
a Lagrange multiplier. The BRST transformations (\ref{s})
represent a trivial BRST algebra for the BRST doublet
$(a^i, c^i)$.
 By an obvious mirror symmetry to the BRST
transformations (\ref{s}), we demand the following anti-BRST
transformations hold:
\begin{equation}
                                                       \label{bars}
\bar s a^{i}= \omega^{ji}\bar c_{j},   \quad
\bar s \bar c_{i}= 0,                    \quad
\bar s c^{i}=  i\omega^{ij}q_{j},          \quad
\bar s q_{i}= 0,
\end{equation}
 The definition (\ref{bars}) implies $\bar s^{2} = 0$,
and it can be easily checked that the BRST and anti-BRST
operators anticommute,
\begin{equation}
                                                         \label{ss}
s\bar s + \bar s s = 0.
\end{equation}
 By definition, $s$ and $\bar s$ anticommute with
$d_t=dt\partial_t$, so that $(d_t+s+\bar s)^2=0$.

 To construct BRST invariant Lagrangian one proceeds as follows.
The partition function (\ref{Z0}) becomes
\begin{equation}
                                                          \label{Z}
Z = \int D[X]\, \exp iI,
\end{equation}
where the measure $D[X]$ represents the path integral over the fields
$a$, $q$, $c$, and $\bar c$.
 The total action $I$ is the trivial action $I_0$ plus $s$-exact part,
\begin{equation}
                                                          \label{I}
I = I_0 + \int dt \, sB.
\end{equation}
 Since $s$ is nilpotent, $I$ is BRST invariant for any choice of $B$,
with $sB$ having ghost number zero.
 Since $\omega_{ij}$ is antisymmetric, all terms quadratic in fields
are identically zero, and we choose judiciously the "gauge-fermion"
$B$ to be linear in the fields.
 The form of $B$ is typical, namely,
(antighost)$\times$(gauge fixing condition),
\begin{equation}
                                                           \label{B}
B = i\bar c_{i}(\partial_{t}a^{i} - h^{i}
              - \alpha a^{i} - \gamma \omega^{ij}q_{j}).
\end{equation}
where $\alpha$ and $\gamma$ are parameters.
 Applying the BRST operator we find that the first two terms in
(\ref{B}) give rise to the term of the form
(Lagrange multiplier)$\times$(gauge fixing condition) and the ghost
dependent part,
\begin{equation}
                                                          \label{sB}
sB = q_{i}(\partial_{t}a^{i} - h^{i})
     + i\bar c_{i}(\partial_{t}c^{i} - sh^{i})
     -  \alpha (q_i a^i - i\bar c_{i}c^{i}).
\end{equation}
 As a feature of the theory under consideration, the
$\gamma$ dependent term vanishes because $\omega_{ij}$ is
antisymmetric.
 To find $sh^i$, we note that
$\delta h^{i}= (\partial_{k}h^{i})\delta a^{k}$,
and hence $sh^i = c^{k}\partial_{k}h^{i}$.
 Thus, the resulting Lagrangian becomes
${\cal L} = {\cal L}_{0} + {\cal L}_{gf} +
            {\cal L}_{gh}+ {\cal L}_{\alpha}$,
\begin{equation}
                                                         \label{L}
{\cal L} =  {\cal L}_{0}
  +  q_{i}(\partial_{t}a^{i} - h^{i})
  + i\bar c_{i}(\partial_{t}\delta^{i}_{k} - \partial_{k}h^{i})c^{k}
  - \alpha (q_i a^i - i\bar c_{i}c^{i}).
\end{equation}
 The total Lagrangian ${\cal L}$ is BRST invariant by construction,
$s {\cal L} = 0$.

 In the delta function gauge, {\sl i.e.} at $\alpha = 0$, it
reproduces exactly, up to ${\cal L}_{0}$, the Lagrangian, which has
been derived in \cite{GR-89}-\cite{GR-90a}, in the
path integral approach to Hamiltonian mechanics by the Faddeev-Popov
method.
 Indeed, by integrating out the fields $q$, $c$ and $\bar c$ we
obtain from (\ref{Z}) the partition function in the form
\begin{equation}
                                                     \label{ZGozzi}
Z = \int Da\, \delta(a - a_{cl})\exp iI_{0},
\end{equation}
where $a^{i}_{cl}$ denote solutions of Hamilton's equation,
$\partial_{t}a^{i}=h^{i}$.
 The partition function (\ref{ZGozzi}), with $I_{0}=0$, has been
used in \cite{GR-89} as a starting point of the
path integral approach to classical mechanics.

 The delta function constraint in (\ref{ZGozzi}) corresponds,
evidently, to the Faddeev-Popov gauge fixing condition and leads to
integration over all paths with a delta function concentrating
around the integral trajectories of the Hamiltonian flow.
 Since the underlying Lagrangian is zero the theory (\ref{L}) is
defined only by the gauge fixing term.
 Thus, the partition function (\ref{Z}), with appropriate boundary
conditions, represents {\sl one-dimensional cohomological field
theory} describing Hamiltonian systems\cite{Aringazin-94-HJ}.

 We note that the $\alpha$ dependent terms in the Lagrangian
(\ref{L}) can be absorbed by redefining time derivative by the
shift, $\partial_t \to \partial_t-\alpha$.
 Notice that the latter form is strongly reminiscent of a gauge
covariant derivative.

 The way to construct explicitly BRST and anti-BRST invariant
Lagrangian is to use both $s$ and $\bar s$ operators, namely,
${\cal L'} = {\cal L}_0 + s\bar s B'$, with $B'$ being of ghost
number zero.
 With the choice
\begin{equation}
                                                       \label{B'}
B' =i\omega_{ik}a^i(\partial_t a^k - h^k),
\end{equation}
we find the Lagrangian in the form
\begin{equation}
                                                       \label{L'}
{\cal L}' =  {\cal L}_{0}
   +  q_{i}\partial_ta^{i} - a^{i}\partial_t q_{i}
   + i(\bar c_{i}\partial_t c^{i} +  c_{i}\partial_t \bar c^{i})
   - q_{i}h^{i}
   + i\bar c_{i}\partial_{k}h^{i}c^{k}.
\end{equation}
We observe that the fields appear in a more symmetric way compared
to (\ref{L}).
 We will see in Sec.~\ref{twist} that this form of the Lagrangian
arises in a superfield treatment of the theory.
 It is straightforward to check that the two Lagrangians, ${\cal L}$,
at $\alpha = 0$, and ${\cal L}'$, differ by the derivative term,
\begin{equation}
                                                       \label{LL'}
{\cal L}'={\cal L}-\frac{1}{2}\partial_t(a^iq_i-ic^i\bar c_i),
\end{equation}
implying thus the same equations of motion.
 Also, we conclude that there is no room for the $\alpha$ dependent
terms, in this $s\bar s$ construction, so that in contrast to
(\ref{L}) the Lagrangian (\ref{L'}) is invariant under the following
$Z_2$ symmetry:
\begin{equation}
                                                        \label{Z2}
t\to -t,      \quad
a^i\to a^i,     \quad
q_i\to -q_i,      \quad
c^i\to c^i,         \quad
\bar c_i\to-\bar c_i, \quad
h^i\to -h^i.
\end{equation}
 Coupling the system to the "gauge field" $\alpha$ spoils this
symmetry.

 In Table~\ref{table} we collect the fields of the cohomological
classical mechanics for the reader convenience
(see also \cite{GRT-89}).
\begin{table}
\centering
\begin{tabular}{|c|c|c|c|}
\cline{1-4}
Field &Meaning             &Geometrical meaning &Ghost number\\
\cline{1-4}
$a$ &map $M^1\to M^{2n}    $ &coordinates on $M^{2n}$      &0\\
\cline{1-4}
$c$ &sdiffeomorphism ghost & differential $da^i$           &+1\\
\cline{1-4}
$\bar c$&anti-ghost of $c$ & $\partial /\partial c^i$      &-1\\
\cline{1-4}
$q$ &Lagrange multiplier   & $-i\partial /\partial a^i$    &0\\
\cline{1-4}
$h$ &vector field          &symplectic vector field        &0\\
\cline{1-4}
\end{tabular}
\caption{The fields of cohomological classical mechanics.}
\label{table}
\end{table}

 The (anti-)BRST symmetry is, in fact, an inhomogeneous part of
larger symmetry of the Lagrangian (\ref{L'}), namely, inhomogeneous
symplectic ISp(2) group symmetry, which is generated by the
following charges\cite{GR-89}:
\begin{eqnarray}
                                                     \label{QCK}
     Q = ic^{i}q_{i},                                \quad
\bar Q = i\bar c_{i}\omega^{ij}q_{j},                \quad
     C = c^{i}\bar c_{i},                            \\
     K = \frac{1}{2}\omega_{ij}c^{i}c^{j},           \quad
\bar K = \frac{1}{2}\omega^{ij}\bar c_{i}\bar c_{j}. \nonumber
\end{eqnarray}
 Here, $Q$ and $\bar Q$ are the BRST and anti-BRST charges
respectively; $sb^i=[Q,b^i]$ and $\bar s b^i=[\bar Q,b^i]$ for a
generic field $b^i$.
 The generators (\ref{QCK}) form the algebra of ISp(2) group,
\begin{eqnarray}
                                                       \label{ISP2}
\bigl[Q,Q\bigr] =
\bigl[\bar Q, \bar Q\bigr] =
\bigl[Q, \bar Q\bigr] = 0,                             \nonumber\\
\bigl[C,Q\bigr] = Q,                                   \quad
\bigl[C, \bar Q\bigr] = - \bar Q,                      \nonumber\\
\bigl[K, Q] = \bigl[\bar K,\bar Q\bigr] = 0,           \nonumber\\
\bigl[\bar K, Q\bigr] = \bar Q,                        \quad
\bigl[K, \bar Q\bigr] = Q,                                      \\
\bigl[K, \bar K\bigr] = C,                             \nonumber\\
\bigl[C,K\bigr] = 2K,                                  \quad
\bigl[C,\bar K\bigr]= -2\bar K.                        \nonumber
\end{eqnarray}
 As it has been found\cite{GRT-89}-\cite{GRT-92} the ISp(2) algebra
(\ref{ISP2})
reflects the full machinery of the Cartan calculus on symplectic
manifold $M^{2n}$, with the correspondences given in
Table~\ref{table}.
 The (anti-)BRST operator $Q$ ($\bar Q$) is naturally associated
with an exterior (co-)derivative $d$ ($d^*$) on $M^{2n}$.

 The ghosts $c^{i}$ form a basis for the tangent space $TM^{2n}$
and act by exterior multiplication on the cotangent space
$T^{*}M^{2n}$, for which the anti-ghosts $\bar c_i$ form the basis
dual to $c^{i}$.
 They fulfill the Dirac algebra,
\begin{equation}
                                                         \label{cc}
\{ c^i, c^j \} = \{ \bar c_i, \bar c_j \} = 0, \quad
\{ c^i, \bar c_j \} = \delta^i_j,
\end{equation}
with the equal time anticommutators, and can be treated as the
creation and annihilation operators acting on a Fock space.
 The basic field $a^i$ and Lagrange multiplier $q_i$ satisfy the
following commutation relation:
\begin{equation}
                                                         \label{aq}
[a^i, q_j] =i\delta^i_j,
\end{equation}
while the other equal time commutators between all the fields are
identically zero.

 Hamilton function, ${\cal H}$, associated with the Lagrangian
(\ref{L}) can be readily derived,
\begin{equation}
                                                          \label{H}
{\cal H} =  q_{i}h^{i} + i\bar c_{i}c^{k}\partial_{k}h^{i}
         + \alpha a^{i}q_{i} + i\alpha C,
\end{equation}
 In the ghost-free part, it covers, at $\alpha = 0$,
the usual Liouvillian, $L=-h^{i}\partial_{i}$, of ordinary classical
mechanics derived in the operator formulation of classical mechanics
by Koopman and von Neumann\cite{Koopman-Neumann}.
 The Hamilton function ${\cal H}$ is a generalization of the
Liouvillian to describe evolution of the form valued probability
distribution, $i\partial_{t}\rho(a,c) = {\cal H}\rho(a,c)$,
instead of the usual distribution function (zero-form) governed by
the Liouville equation, $\partial_{t}\rho(a) = - L\rho(a)$.
 Here, we mean $\rho(a,c)$ is expanded in anticommuting variables
$c^i$ giving a set of $p$-ghost terms corresponding to a set of
$p$-forms, $0\leq p \leq 2n$, on $M^{2n}$.
 From a geometrical point of view, it is highly remarkable that
${\cal H}$, at $\alpha=0$, is proportional to a Lie derivative
along the Hamiltonian vector field, ${\cal H}=-i\ell_h$, applied
this time to $p$-forms\cite{GRT-89,GRT-92}.
 As we will see in Sec.~\ref{twist}, solving the generalized
Liouville equation, for stationary form valued distributions,
is equivalent to solving a cohomology problem.

 It is worthwhile to note here that the $p$-form states and
observables arise naturally also in the supersymmetric quantum
mechanics\cite{Witten-82-JDG,Witten-82-NP} and in the
(Landau-Ginzburg) $N=2$ supersymmetric
models\cite{Vafa-91,Cecotti-92,Cecotti-91}, which in various
aspects will serve, in Sec.~\ref{twist}, as a guide line for
dealing with such an extension.
 Also, we note that the properties of the ground states in the
two-dimensional topological models are studied\cite{Cecotti-91}
also via dimensional reduction to one-dimensional models.

 Besides symplectic structure of the target space, the model
(\ref{L'}) reveals symplectic structure provided by the field
content.
 Recall first that the phase space $M^{2n}$ is usually
considered as a cotangent bundle over configuration space $M^n$,
$x^{\alpha}\in M^n$.
 Analogously, let us consider cotangent bundle $M^{4n}$ over
$M^{2n}$ endowed by symplectic two-form $\tilde\Omega$ with local
coordinates $\tilde y^c=(q_i,a^j)$; $c,d=1,\dots,4n$.
 Here, $q_i$ and $a^j$ are canonical conjugates which is indicated
by (\ref{aq}).
 $M^{4n}$ can be thus viewed as the second generation phase space
with a base space $M^{2n}$ and natural projection
$p:\, M^{4n}\to M^{2n}$ provided by $(q_i,a^j)\mapsto (0,a^j)$.
 Enlarging the bundle $M^{4n}$ by a Grassmannian part with
coordinates $(c^i,\bar c_j)$, we define the cotangent superbundle
$M^{4n|4n}$ equipped by block diagonal supersymplectic matrix
$||\Omega_{ab}||=\mbox{diag}(||\tilde\Omega_{cd}||,||E_{cd}||)$,
where $E$ is $4n\times 4n$ unit matrix, with local coordinates
$y^a=(\tilde y^d,c^i,\bar c_j)$; $a,b=1,\dots,8n$.
 The graded Poisson brackets in $M^{4n|4n}$ has been introduced in
\cite{GRT-89,GR-92d}, and can be defined by using $\Omega$
in a standard way:
\begin{equation}
                                                        \label{epb}
\{F,G\}_{\Omega}=(\partial_aF)\Omega^{ab}(\partial_bG),
\end{equation}
where $F$ and $G$ are functions on $M^{4n|4n}$ and
$\partial_a=\partial/\partial y^a$;
$\Omega_{aa'}\Omega^{a'b}=\delta_a^b$.

 Thus, the Hamiltonian (\ref{H}) and charges in (\ref{QCK}) can
be treated as functions on $M^{4n|4n}$ acting by taking the graded
Poisson brackets.
 It is a matter of straightforward calculations to verify that the
ISp(2) algebra relations (\ref{ISP2}) hold, with the graded
commutators replaced by the graded Poisson brackets;
for example, $\{\bar K,Q\}_{\Omega}=\bar Q$.
 We note that this is, in fact, a nontrivial result because of
emerging of the supersymplectic structure $\Omega$ having no
counterpart in the Cartan calculus.
 The probability distribution forms $\rho$ on $M^{2n}$ can be viewed
in general as functions on $M^{4n|4n}$, with the graded Poisson
bracket algebra being the algebra of classical observables.
 The Hamilton function (\ref{H}), at $\alpha=0$, is ad$_H$ operator
acting on functions on $M^{4n|4n}$ and represents horizontal vector
field in the fiber bundle. The Schrodinger-like equation for
evolution of the distribution form can be rewritten in
a Hamiltonian form, $i\partial_t\rho=\{{\cal H},\rho\}_\Omega$.
 In a sense, we can say that the model is twice symplectic.

  Due to the structure of the Hamilton function (\ref{H})
the partition function (\ref{Z}) can be factorized into three
different sectors: The Liouvillian sector, the ghost sector,
and the $\alpha$ dependent sector.

 In the following, we use the delta function gauge omitting
the $\alpha$ dependent terms in (\ref{H}), one of which is
the ghost number operator $C$.

\subsection{SLAVNOV IDENTITY}\label{Slvnv}

  In order to draw further parallels with the topological quantum
field theories, the point of an immediate interest is to translate
the BRST invariance of the theory under consideration into Slavnov
identity.
 Particularly, the Slavnov identity technique is
used\cite{Oliveira-92} to study anomalies and renormalizability of
a theory and to incorporate all the symmetries and constraints of
a model (BRST invariance, vector supersymmetry, ghost equations,
etc.).
 Since the theory under consideration is linear and one-dimensional,
its perturbative properties and anomalies can be reliably derived
from general arguments. It is instructive, however, to prove
explicitly that it is indeed perturbatively trivial and symplectic
diffeomorphism anomaly free.

 In order to write down the Slavnov identity, we introduce a set of
invariant external sources $(J^a, J^q, J^c, J^{{\bar c}})$ coupled
to the BRST variations of the fields,
\begin{equation}
                                               \label{Iext}
I_{ext}
= \int dt (J^a sa + J^q sq + J^c sc + J^{{\bar c}}s{\bar c}).
\end{equation}
 According to (\ref{s}) the total action,
\begin{equation}
                                                \label{Sigma}
\Sigma = I + I_{ext}
= \int dt
(      q_i(\partial_t a^i-h^i)
+ \bar c_i \partial_t c^i
+ \bar c_i \partial_j h^i c^j
+ J^a_ic^i + iJ_{\bar c}^iq_i),
\end{equation}
does not depend on $J^c$ and $J^q$ since $sc=sq=0$, while the other
BRST transformations in (\ref{Iext}) are linear.
This linearity implies that there are no "radiative corrections"
to these transformations so that linear dependence of the action
(\ref{Sigma}) on the BRST sources is radiatively preserved.

 It is straightforward to check that the extended action
(\ref{Sigma}) satisfies the following Slavnov identity:
\begin{equation}
                                                 \label{Slavnov}
{\cal S}(\Sigma) = 0,
\end{equation}
where
\begin{equation}
                                                  \label{SSigma}
{\cal S}(\Sigma) =\int dt
\Bigl(
\frac{\delta \Sigma}{\delta J^a_i}
\frac{\delta \Sigma}{\delta   a^i}
+
\frac{\delta \Sigma}{\delta J_{\bar c}^i}
\frac{\delta \Sigma}{\delta    \bar c_i }
\Bigr).
\end{equation}
 The corresponding extended BRST operator $B_{\Sigma}$ is linear,
\begin{equation}
                                                  \label{BSigma}
B_{\Sigma} =\int dt
\Bigl( c^i\frac{\delta}{\delta a^i}
      +iq_i\frac{\delta}{\delta\bar c_i }
\Bigr),
\end{equation}
and provides no extension to the BRST sources.
 It is easy to verify that $B_{\Sigma}$ is nilpotent,
$B_{\Sigma}^2=0$.
 With the absence of the radiative corrections to this equality
we arrive at the conclusion that there are no "quantum deformations"
(no surprise certainly).

 In topological Yang-Mills field theories, nontrivial cohomology
of an extended BRST operator in the space of integrated
polynomials in fields and BRST sources is referred to as a gauge
anomaly\cite{Oliveira-92}. So, anomaly may come from nontrivial
cohomology of the extended BRST operator $B_{\Sigma}$ in such a space.
 However, it is easy to verify that its cohomology is trivial since
the fields in (\ref{BSigma}) appear only in BRST doublets.
 Note that one should take into account all symmetries of
the theory to write down extended BRST operator.
 Since in our case there are no additional symmetries to be
incorporated, this completes the prove that the Slavnov identity
is symplectic diffeomorphism anomaly free.

\section{PHYSICAL STATES AND TOPOLOGICAL TWIST}\label{twist}

 Due to the BRST and anti-BRST invariance of the theory we will
study in this Section the BRST and anti-BRST invariant states.
 The possible physical states, $\rho \equiv |$phys$\rangle$,
are then found as solutions of the system of equations consisting
of the BRST and anti-BRST cohomology
equations\cite{GRT-91c,Aringazin-93-HJ,Aringazin-93-PL},
\begin{equation}
                                                 \label{Qrho}
     Q \rho = 0, \quad
\bar Q \rho = 0.
\end{equation}
They are equivalence classes of appropriate $Q$ and $\bar Q$
cohomologies, $\rho \sim \rho + Q\rho' +\bar Q\rho''$.

\subsection{$d=1$, $N=2$ SUPERSYMMETRIC MODEL}\label{Ssym}

 To study the physical states, we exploit the identification of
the twisted (anti-)BRST operator algebra with $N=2$ supersymmetry
which is usually performed in topological quantum field
theories\cite{Witten-82-JDG}.
 Conventionally, the topological twist is used to obtain BRST theory
from a supersymmetric
one\cite{Witten-88,Witten-82-JDG,Anselmi-92,Witten-92}.
 Below we use the twist to obtain, conversely, supersymmetry from
the BRST and anti-BRST symmetries.

 Using the twisted BRST and anti-BRST operators
\begin{equation}
                                                       \label{Qbeta}
Q_{\beta} = e^{\beta H}Qe^{-\beta H}
          = Q - \beta c^i \partial_i H, \quad
\bar Q_{\beta} = e^{-\beta H} \bar Q e^{\beta H}
               = \bar Q + \beta \bar c_i\omega^{ij}\partial_j H,
\end{equation}
where $\beta \geq 0$ is a real parameter, one can easily find that
they are conserved nilpotent supercharges and their anticommutator
closes on the Hamilton function,
\begin{eqnarray}
                                                    \label{susy}
\{ Q_{\beta},Q_{\beta}\}
=\{ {\bar Q}_{\beta},{\bar Q}_{\beta}\}=0,          \nonumber \\
\{ Q_{\beta},{\cal H}\}
= \{ {\bar Q}_{\beta},{\cal H}\}=0,                 \\
\{ Q_{\beta},\bar Q_{\beta}\} = 2i\beta {\cal H}.   \nonumber
\end{eqnarray}
 Consequently, these supercharges, together with the Hamilton
function ${\cal H}$, build up global $N=2$ supersymmetry.
 The supersymmetry transformations leaving the Hamilton function
${\cal H}$ invariant are
\begin{equation}
                                                    \label{susytr}
\delta_s a^i = c^i,                                 \quad
\delta_s c^i = 0,                                   \quad
\delta_s \bar c_i = iq_i - \beta \partial_i H,      \quad
\delta_s q_i = -i\beta c^k\partial_i\partial_k H,
\end{equation}
\begin{equation}
                                                    \nonumber
\bar\delta_s a^i = \omega^{ki}\bar c_k,             \quad
\bar\delta_s c^i = i\omega^{ik}q_k
                 + \beta\omega^{ik}\partial_k H,    \quad
\bar\delta_s \bar c_i = 0,                          \quad
\bar\delta_s q_i =
          i\beta\omega^{jk}\bar c_j\partial_i\partial_k H.
\end{equation}
 It is important to note that the topological twist (\ref{Qbeta})
does not change the Lagrangian of the theory.
 It is well known that in the case of $d \geq 2$ topological
field theories a crucial property of the twist is that it changes
statistics of some fields.
 Apart from the case of the $d \geq 2$ theories, we are dealing with
the fields which have no spin because it makes no sense in $d=1$
case. So, the problem of changing of statistics is irrelevant.

 The supersymmetry (\ref{susy}) has been originally found by Gozzi
and Reuter\cite{GR-90}, who stressed that it is of
fundamental character in Hamiltonian systems.
 Particularly, this supersymmetry has been used\cite{GR-90} to
derive the classical Kubo-Martin-Schwinger condition justifying
algebraically the preference of the Gibbs distribution and has been
related\cite{GRT-91a,GRT-92,GRT-91c} to the regular/nonregular motion
regimes in Hamiltonian systems with Hamiltonian, which does not
explicitly depend on time.

 In general, the supersymmetry garantees that there will be a set
of exactly degenerate ground states.
 More specifically, if the supersymmetry is exact the Hamiltonian
system is in the nonregular motion regime since, in this case,
there is only one conserved entity (energy) while if the system
is in regular motion regime, {\sl i.e.} there is at least one
additional nontrivial integral of motion, the supersymmetry is
always broken.
 So, the condition of the supersymmetry breaking is of much
importance and it can be thought of as a criterion to distinguish
between the regimes. We will see shortly that it is naturally
related to the Witten index.
 However, there is a subtlety to make one-to-one correspondence.
Namely, broken supersymmetry does not necessary imply regular
motion regime and, also, nonregular motion regime does not rule
out broken supersymmetry. It seems that the problem lies precisely
in possible degeneracy of the ground state, that is,
\mbox{ker}$\,{\cal H}$ modulo cohomological equivalence may consist
of several elements\footnote{The space \mbox{ker}$\,{\cal H}$ is
finite dimensional since ${\cal H}$ is identified with an elliptic
operator.}.
 The indication of this is that an ergodic Hamiltonian
system is characterized just by {\sl nondegenerate} zero eigenvalue
solution of the Liouville equation. So, one is led to study physics
coming from the degenerate vacuum.
 In the following, we assume there is a discrete set of
supersymmetric ground states. This corresponds to the case of
an elliptic operator on compact manifold.

 The relation of the BRST and anti-BRST symmetries of the original
theory to the $N=2$ supersymmetry (\ref{susy}) is as follows.
 Usually, BRST exact theory is referred to as a topological
theory. Anti-BRST exact theory is viewed as its conjugate,
anti-topological theory.
 The crucial link between these two theories is the
supersymmetric ground state sector,
--referred to as {\sl Ramond sector}-- of the associated
$d=1$, $N=2$ supersymmetric model, in accordance with the
topological-antitopological fusion by Cecotti and
Vafa\cite{Cecotti-91}.
 Namely, the physical states of both the topological theories are
in one-to-one correspondence with the Ramond vacua, as it can be
seen in Sec.~\ref{Chmlg}.

 It is instructive to note here that as it has been argued recently
by Perry and Teo\cite{Perry-92/33}, in the context of topological
Yang-Mills theory, both the BRST symmetry and anti-BRST symmetry
should be taken into account on an equal footing to get a clear
geometrical meaning of the topological theory.
 It is worth stressing that this argument is supported by the
cohomological classical mechanics, in which all the symmetries and
fields have clear geometrical meaning, with both the BRST and
anti-BRST symmetries being incorporated; see Table \ref{table}
and (\ref{ISP2}).
 In fact, this reflects canonical isomorphism between the tangent
and cotangent spaces, $TM^{2n}$ and $T^{*}M^{2n}$, provided by the
symplectic structure.

\subsection{COHOMOLOGY}\label{Chmlg}

 Our next step in studying the physical states is identification of
the $N=2$ supersymmetry (\ref{susy}) with an exterior algebra, in
analogy with the identification made in Witten's supersymmetric
quantum mechanics\cite{Witten-82-JDG}.
 This allows us to relate the supersymmetric properties of the
model to topology of the target space $M^{2n}$.

 To begin with, we mention that it has been argued\cite{GRT-92}
that cohomology of $Q_{\beta}$ is isomorphic to de Rham
cohomology.
 In a strict consideration, to which we are turning now, one
should associate, in a standard way, an elliptic complex to it.
 Namely, we make the following identifications:
$$
d_{\beta}      \leftrightarrow  Q_{\beta},       \quad
\bar d_{\beta} \leftrightarrow  \bar Q_{\beta},
$$
\begin{equation}
                                                         \label{dQ}
\Delta_{\beta}\equiv d_{\beta}\bar d_{\beta}+\bar d_{\beta}d_{\beta}
                     \leftrightarrow
      \{ Q_{\beta},\bar Q_{\beta}\} = 2i\beta {\cal H},
\end{equation}
$$
         (-1)^{p} \leftrightarrow (-1)^{C},
$$
where the exterior (co-)derivative
$d_{\beta} = d+\beta c^i\partial_i H$
($\bar d_{\beta} = d^{*}-\beta{\bar c}_i\omega^{ij}\partial_j H$)
acts on $p$-forms, $\rho \in \Lambda^{p}$, and $C$ is the ghost
number.

 The $d_{\beta}$ cohomology groups,
\begin{equation}
                                                 \label{cohomology}
H^{p}(M^{2n})
=\{ \ker d_{\beta} / \mbox{im}\, d_{\beta}\cap \Lambda^{p} \},
\end{equation}
are finite when $M^{2n}$ is compact.
 According to Hodge theorem, canonical representatives
of the cohomology classes $H^{p}(M^{2n})$ are harmonic $p$-forms,
\begin{equation}
                                                       \label{Drho}
\Delta_{\beta}\rho = 0.
\end{equation}
 They are closed $p$-forms,
\begin{equation}
                                                       \label{drho}
d_{\beta}\rho = 0, \quad \bar d_{\beta}\rho = 0.
\end{equation}
 One can then define $B_{p}$ as the number of independent harmonic
forms, {\sl i.e.}
\begin{equation}
                                                      \label{Betti}
B_{p}(\beta ) = \dim \{ \ker \Delta_{\beta} \cap \Lambda^{p} \}.
\end{equation}
 Formally, $B_{p}$ continuously varies with $\beta$ but, being a
discrete function, it is, in fact, independent on $\beta$ so that
one can find $B_{p}$ by studying the vacua of the Hamilton function
(\ref{H}),
\begin{equation}
                                                       \label{Hrho}
{\cal H}\rho = 0.
\end{equation}
 This equation is, in fact, the only equation we need to study.
Due to (\ref{susy}), it can be rewritten as
\begin{equation}
                                                   \label{Qbetarho}
Q_{\beta}\rho = 0, \quad
\bar Q_{\beta}\rho = 0,
\end{equation}
and defines the Ramond sector. The equivalence between
(\ref{Hrho}) and (\ref{Qbetarho}) needs a comment.
While it is obvious that (\ref{Qbetarho}) implies (\ref{Hrho}),
vice versa may appear to be problematic for spaces with
a lack of a positive-definite scalar product. For example,
when proving that if an external differential form is harmonic then
it is closed and coclosed, one uses the fact that scalar product of
forms is positive definite. So, we define a scalar product
in the space of $p$-forms in a standard way,
$\langle\rho,\rho'\rangle = \int \rho\wedge *\rho'$,
which is positive definite, to ensure that (\ref{Qbetarho})
follows from the harmonic condition (\ref{Hrho}).


 One would expect that the spectrum of the theory is defined by
the whole set of the eigenvalues $\kappa_{i}$ and eigenfunctions
$\rho_{i}$ of the Hamilton function,
${\cal H}\rho_{i} = \kappa_{i}\rho_{i}.$
 However, only the states with zero eigenvalue of ${\cal H}$ are
nontrivial in the BRST and anti-BRST cohomology since
${\cal H}$ commutes with both the BRST and anti-BRST charges.
 Indeed, by a standard argument all the states except for
the ground states are of no relevance due to the supersymmetry
(the superpartners' states give a net zero contribution).
 To be more specific, if we consider deformations $\delta a^{i}$
along the solution of Hamilton's equation, then in order for
$a^{i} + \delta a^{i}$ to still be a solution it has to satisfy the
deformation equation $\partial_{t}\delta a^{i} = \delta h^{i}.$
 This equation is the equation for the conventional Jacobi fields,
first variations, which are tangent to the target manifold,
$\delta a^{i} \in TM^{2n}$, and can be thought of as the "bosonic
zero modes".
 As it is common in topological field theories, these modes are just
compensated by anti-commuting zero modes through the ghost dependent
term, in the Lagrangian (\ref{L'}).
 Indeed, the associated equation of motion for ghosts,
$\partial_t c^i = \partial_k h^i c^k,$ represents BSRT variation of
Hamilton's equation.

 Thus, the physical states are those in the Ramond sector,
{\sl i.e.} satisfying (\ref{Qbetarho}), which
intertwines corresponding BRST and anti-BRST symmetries.
 Topologically, the relation of the supersymmetry equations
(\ref{Qbetarho}) to the cohomology equations (\ref{Qrho}) can be
readily understood by taking into account the fact that the twist
(\ref{Qbeta}) is a homotopy operation.

 According to the identification (\ref{dQ}) with the exterior
algebra, the physical states are harmonic $p$-forms on the target
space $M^{2n}$. In the standard de Rham complex, the $B_{p}$'s are
simply Betti numbers, with the alternating sum,
$\chi = \sum^{2n} (-1)^{p}B_{p}$, being the Euler characteristic
of the symplectic manifold $M^{2n}$.
 Due to the identifications (\ref{dQ}), it is then straightforward
to show that the Witten index\cite{Witten-82-NP}, $\mbox{Tr}(-1)^C$,
is just the Euler characteristic of $M^{2n}$ (cf.\cite{G-93}).
 Here, the conserved charge $C$ is identified with the Fermi number
operator, $F=C$. This is a natural result due to the fact that
the Witten index is completely independent of finite perturbations
of the theory for $N\geq 1$ supersymmetric theories in any
dimensions.
 So, we conclude that the criterion for the regular/nonregular
regimes in Hamiltonian systems which is related to the
supersymmetry breaking is the Witten index.
 To break supersymmetry the Witten index needs to be zero.
 We arrive at the conclusion that the motion regimes are related
to topology of $M^{2n}$.

 Equation (\ref{Hrho}) can be thought of as a generalization of
the {\sl ergodicity} condition equation\cite{Arnold-68},
$L\rho(a) = 0$, of the usual Hamiltonian mechanics which is now
extended to the $p$-ghost ($p$-form valued) distributions,
$\rho = \rho(a,c)$.
 We recall that nondegenerate solution of the latter equation
characterizes an ergodic Hamiltonian system.
 It is wellknown that the general solution of this equation is a
function of Hamiltonian, $\rho = \rho(H(a))$,
 The supersymmetry strengthen this statement by fixing dependence
on $H$.
 Recent studies\cite{GRT-91c,Aringazin-93-HJ,Aringazin-93-PL}
of the physical states (\ref{Qbetarho}) showed that physically
relevant (normalizable) solutions to the generalized ergodicity
equation (\ref{Hrho}) come from the $2n$-ghost sector and have
specifically the form of the Gibbs state characterizing
thermodynamical equilibrium,
\begin{equation}
                                                    \label{Gibbs}
\rho = \kappa K^{n}\exp [-\beta H]
              \leftrightarrow
       \kappa\exp[-\beta H]da^1\wedge\cdots\wedge da^{2n},
\end{equation}
where $\kappa$ is a constant.
 It is important to note that under the field redefinition this
state transforms as $2n$-form rather than as a scalar.  The reason
of this lies in the cohomology.
 The other ghost sectors yield solutions of the form
$\rho = \kappa K^{p}\exp [+\beta H]$, for the even-ghost sectors,
$p = 2, 4, \dots, 2n-2$, and those are either trivial or not
depending on $\beta$, for the odd-ghost sectors.
 It is assumed that $\rho(a,c)$ must be normalizable in each $p$-ghost
sector, {\sl i.e.} $\int\rho(a,da)da^1\wedge\cdots\wedge da^p=1$,
$p=0,\dots,2n$.
 While the result for the even-ghost sectors is reliable and
quite clear, the odd-ghost sector solutions are a bit cumbersome.
 In Appendix, we sketch analysis on solutions in simplest
two-dimensional case\cite{Aringazin-94-HJ},
both to illustrate emerging of the Gibbs distribution and
to clarify the meaning of the odd-ghost sector.

 To summarize, we observe that the supersymmetry is helpful in
obtaining some important results on Hamiltonian systems so that it is
worthwhile to study features of the $d=1$, $N=2$ model more closely.
 We postpone this to Secs.~\ref{Mrs}-\ref{Eqvr}.

 Our next observation is that, in the limit $\beta \rightarrow 0$,
we recover the classical Poincare integral
invariants\cite{Abraham-Arnold} corresponding to $K^{p}$,
$p=1,\dots n$, as the solutions of (\ref{Qbetarho}), which are
indeed invariants under Hamiltonian flow, ${\cal H}K^{p}=0$.
 In particular, the $2n$-ghost $K^{n}$, which we are viewing as
cohomological representative of the unit, corresponds to the
volume form $\omega^n$ of the phase space conservation of which
is statement of the Liuoville theorem.
 They are fundamental BRST invariant (topological) observables of
the theory, $\{Q,K^{p}\}=0$, and form the classical cohomology ring,
$K^{n+1} = 0$.
 Indeed, in the untwisting limit, $\beta \rightarrow 0$,
the supersymmetry generators $Q_{\beta}$ and $\bar Q_{\beta}$
become the original BRST and anti-BRST operators, respectively.
 Geometrically, this follows from $dK^{p}=0$ since
$K^{p} \leftrightarrow \omega^{p}$ and $d\omega = 0$.
 Similarly, the conjugates of the Poincare invariants, $\bar K^{p}$,
are the anti-BRST invariant (anti-topological) observables,
$\{\bar Q, \bar K^{p}\}=0$, $p=1,\dots n$,
which are powers of the Poisson bivector $\bar K$.

 The following comments are in order.

 (i) We recall that the physical states considered above are
defined as the BRST {\sl and} anti-BRST invariant ones.
 We see that this requirement, which is equivalent to unbroken
supersymmetry, put strong limits on the possible
physical states restricting it in effect to the
(highest) $2n$-ghost sector (\ref{Gibbs}).
 This is, in fact, a physically acceptable result leading to
non-zero expectation values of scalar (ghost-free) observables,
$\langle A(a)\rangle=\int A\rho$, whereas for the other
$p$-ghost ($p$-form valued) observables, $1<p\leq 2n$, we have that
their averages are identically zero.
 However, we should note that one can consider only the BRST
invariant theory, as the topological sector of the $d=1$, $N=2$
supersymmetric model.

 (ii) Since the vacuum distribution forms, $\rho$, are annihilated
by the supersymmetry charges (\ref{Qbeta}) the modified forms
$\lambda=\exp[-\beta H]\rho$ and $\tilde\lambda=\exp[\beta H]*\rho$
are $d$-closed. Here, $*$ denotes Hodge duality operator and
according to even-dimensionality of $M^{2n}$ we have $d=*d^**$ and
$d^*=-*d*$.
 The forms $\tilde\lambda$ can be viewed as representatives of the
{\sl relative} de Rham classes (see, for example,
Ref.\cite{Cecotti-91}), $H^p(M^{2n},D)$, with $D \subset M^{2n}$
being the region where $\beta H$ is greater than a certain large
value.
 The forms $\lambda$ correspond to the dual cohomology of the
associated cycles, which form an integral basis for the Ramond
vacua.
 A remarkable feature of the relative de Rham cohomology is that it
can be nontrivial even if the usual de Rham classes of $M^{2n}$
are trivial; for example, when $M^{2n} = R^{2n}$.
 However, we will not discuss further on the relative cohomology
here restricting consideration on compact $M^{2n}$, for which case
the usual de Rham classes are nontrivial.

\subsection{CONNECTION TO MORSE THEORY}\label{Mrs}

 Let us to note that there appears to be no relation of the $d=1$,
$N=2$ model (\ref{susy}) to Morse
theory\cite{Milnor-73,Bott-80,Labastida-88}
quite analogous to that found in Witten's supersymmetric quantum
mechanics\cite{Witten-82-JDG}, because of the absence of the term
quadratic in $h^{i}$, in the Hamilton function (\ref{H}), which
would play a role of the potential energy.

 An immediate reason is that the symplectic two-form $\omega$
is {\sl closed} so that this does not allow us to construct, or
obtain by the BRST procedure, non-vanishing terms in the Hamilton
function quadratic in fields, except for the ghost-antighost term,
which contains the Hessian, $\partial_i \partial_j H(a)$.
 Put differently, this is due to the symplectic structure of the
target space $M^{2n}$ which has been used as the {\sl only}
differential geometry structure to construct the cohomological
classical mechanics.
 On the other hand, quadratic term, which is natural in
(topological) quantum field theories when one uses Riemannian
(or Kahler) structure of the target space, would produce stochastic
contribution (Gaussian noise) to the equations of motion\cite{PS-79}
that would, clearly, spoil the deterministic character of the
Hamiltonian mechanics.
 As a consequence, we can think of linearity of the Lagrangian
(\ref{L'}) in the {\sl commuting} fields as a condition of the
classical deterministic behavior of the system\footnote{Liouvillian
$L$ is a linear differential operator.}.
 It should be emphasized here that the path integral approach to
classical mechanics\cite{GRT-89} relied basically on the work by
Parisi and Sourlas\cite{PS-79} who studied classical stochastic
equations.

 Due to the absence of a term quadratic in $h^i$ in the Hamilton
function (\ref{H}), there are no localized states and solitons
similar to that of supersymmetric quantum mechanics
which could be used to find a deeper connection between the
(twisted) $d=1$, $N=2$ model and Morse theory.
 Although the Hamilton function (\ref{H}) is linear in the
commuting fields, it contains the Hessian of the Hamiltonian
$H(a)$, which can serve as Morse function, the number of
isolated critical points of which are known to be related to
Euler characteristic. This link has been analyzed in detail in
\cite{GR-90a}.
 The critical points here are simply stationary points of the
Hamiltonian flow, $h^i = 0$, with the number of critical points
\begin{equation}
                                                    \label{crit}
\sum\limits_{dH=0}{} sign (\det ||\partial_i \partial_j H(a)||).
\end{equation}
 We will mention a bit more on the connection to Morse theory
in Sec.~\ref{correlation}, in the context of partition function.

\subsection{LANDAU-GINZBURG FORMULATION}\label{LG}

 As a preliminary observation, we note that the form of the
definition (\ref{Qbeta}) suggests that the Hamiltonian $H$
plays the role analogous to that of superpotential in
supersymmetric quantum mechanics (one-dimensional version of
the Landau-Ginzburg model).
 Namely, action of the supercharge $Q_{\beta}$ on forms can
be casted in the form
$Q_{\beta}\rho = \partial\rho + dH\wedge\rho$,
where we have rescaled $H$ by $\beta$ for a moment.

 Due to the underlying $N=2$ supersymmetry (\ref{susy}),
it is instructive to give a superfield representation of the
cohomological classical mechanics which is usually used in the $N=2$
supersymmetric models as well as in the topological Yang-Mills
theory\cite{Perry-92/33}, to write down the basic settings in a
simple closed form.
 Advantage of this formulation is that one could readily see
whether the cohomological classical mechanics admits
a kind of Landau-Ginzburg description\cite{Witten-88a,Vafa-91}
so that it could be largely understood in terms of superpotential.
 Since the superpotential is, in effect, the ordinary classical
Hamiltonian $H(a)$, this arises to possibility to classify
integrable Hamiltonian systems, within the Landau-Ginzburg
framework.

 Collection of fields composing a Landau-Ginzburg type system
is the following: real field $a^i$, two anticommuting real fields
$c^i$ and $\bar c_i$, real field $q^i$, and superpotential $W$
solely responsible for the "interaction" terms.
 We choose a single superfield as follows (cf.\cite{GRT-89}):
\begin{equation}
                                                     \label{sfield}
X^i(t,\theta_1,\theta_2)
= a^i(t) - i\theta_1c^i(t) + i\theta_2\omega^{ij}\bar c_j(t)
          +i\theta_1\theta_2 \omega^{ij}q_j(t),
\end{equation}
where $\theta_I$,  $I=1,2$, are real anticommuting parameters,
and the component fields are nothing but a collection of the BRST
doublets.
 Geometrically, the components of the superfield form local
coordinates of the tangent and cotangent fiber bundles over the
phase space $M^{2n}$ (see Table \ref{table}).

A manifestly covariant Lagrangian
\begin{equation}
                                                     \label{Lsuper}
{\cal L} = \int d\theta_1 d\theta_2 \,
\Bigl\{\frac{1}{2}\omega_{ij}X^iD_1D_2 X^j +iW(X)\Bigr\},
\end{equation}
can be written with the help of the covariant derivatives in the
superspace with local coordinates $(t,\theta_1,\theta_2)$,
\begin{equation}
                                                        \label{DI}
D_1 = \frac{\partial}{\partial\theta_1}
      + i\theta_2\frac{\partial}{\partial t},
\quad
D_2 = \frac{\partial}{\partial\theta_2}
      + i\theta_1\frac{\partial}{\partial t},
\end{equation}
$D_1^2=D_2^2=0$, $\{D_1,D_2\}=2i\partial_t$, and the superpotential
$W(X)$, which is a real analytic function of $X$.
 In terms of the component fields, we find the Lagrangian
(\ref{Lsuper}) in the form
\begin{eqnarray}
                                                      \label{Lcomp}
{\cal L} = q_i\partial_t a^i - a^i\partial_t q_i
 +i(\bar c_i\partial_t c^i + c^i\partial_t \bar c_i ) \\
                                                       \nonumber
+ \frac{\partial W(a)}{\partial a^i}\omega^{ij}q_j
- i\omega^{ij}\bar c_i
  \frac{\partial^2 W(a)}{\partial a^k\partial a^j}c^k,
\end{eqnarray}
where $W(a)$ is the lowest component of the superpotential.
 With the identification $W(a) = H(a)$, the Lagrangian (\ref{Lcomp})
covers the original Lagrangian (\ref{L'}).
 So, we conclude that one can start with the one-dimensional
Landau-Ginzburg $N=2$ model (\ref{Lsuper}) and obtain via
topological twist the cohomological theory (\ref{L'}) with already
gauge fixing. Here, topological twist provides transition from
supercharges to BRST charges. Note that the Lagrangian (\ref{Lsuper})
provides the action to be of the form of a D-term.

 For completeness, let us to note that the whole set of BRST and
anti-BRST transformations (\ref{s}) and (\ref{bars}) takes the
form of the following constraint (see also \cite{GRT-89}):
\begin{equation}
                                                    \label{stheta}
(s_{I}-\frac{\partial}{\partial\theta_I})X^i = 0,
\end{equation}
where we have denoted $s_1=s$, $s_2=\bar s$, saying that
the BRST and anti-BRST operators can be treated as derivatives
in the odd coordinates of superspace.
 Therefore, the topological invariance of the action is obvious in
superspace because of supertranslation invariance,
$X^i(t,\theta_1,\theta_2) \to
X^i(t,\theta_1+\theta_1',\theta_2+\theta_2')$,
of the Berezin integration.

 It is simple but important consequence of supersymmetry algebra
that the action with the Lagrangian (\ref{Lsuper}),
like any D-term since it is the highest component, can be written
{\sl both} as $I=\{Q_{\beta}, \xi\}$ and
$I=\{\bar Q_{\beta},\bar\xi\}$, where $\xi$ is some
odd field integrated over time.
 Note also that we were not forced to use a F-term, which is defined
as an integral over only half of superspace, to reproduce the
original action.

 The above are the essential ingredients necessary for
arguments in analyzing implications of the supersymmetric
structure of cohomological Hamiltonian mechanics.

 For example, it is wellknown that the symplectic two-form $\omega$
can not be in general defined to be constant globally on compact
$M^{2n}$, so the question arises as to cohomology classes of
$\omega$ in $M^{2n}$.
 Two choices of the Lagrangian, for which $\omega$ are in
different cohomology classes, differ by F-term.
 On the other hand, since the action is of the form of a D-term
we have no topological effect of changing the classes which could
be in principle considered by perturbing the action by a F-term.

 From such a general point of view, it may seem that the problem
on cohomological class $[\omega]$ of $\omega$ in the theory is
extrinsic. However, in fact it has a direct link to time
reparametrization invariance of the model.

 Consider the time reparametrization $t \to e^\phi t$, where
$\phi$ is a parameter ($d=1$ Lorentz transformation).
 In the Lagrangian (\ref{Lsuper}), the only effect it produces
is the scaling $\omega\to e^{-\phi}\omega$.
 This implies scaling of the volume of compact phase space,
$V=\int_{M^{2n}}\omega^n \to e^{-n\phi}V$, so that with the factor
$1/V^{1/n}$ the Lagrangian (\ref{Lsuper}) becomes time
reparametrization invariant.
 Without loss of generality, assume that $\omega$ is an exact
two-form in some region $U\subset M^{2n}$, that is,
$\omega=d\vartheta$, where $\vartheta=\vartheta_ida^i$ is a
symplectic one-form.
 The general consistency requirement is then that the Wilson loop
integral $\exp[2\pi i\oint_{\partial D}\vartheta]$ should not depend on
the disk $D$, for which $\partial D$ is its boundary.
 Hence, we must have $\int_{S^2}\omega=k$, where $k$ is an
integer number.
 The scaling of $\omega$ demands $k=0$, so we arrive at
the conclusion that unless the condition
\begin{equation}
                                                  \label{intomega}
\int\limits_{S^2}\omega = 0, \qquad \forall S^2\subset M^{2n},
\end{equation}
is satisfied, the time reparametrization invariance of the model
is broken.

 The condition (\ref{intomega}) is known in mathematical
literature\cite{Niemi-94,Floer-86} in another aspect,
and essentially implies that the fundamental homotopy group
must be nontrivial, $\pi_1(M^{2n})\not=0$.
 This can be seen as follows.
 As the symplectic two-form is closed but in general is not
exact, the condition (\ref{intomega}) means that all cycles
$S^2\subset M^{2n}$ are contractible, $[\omega]\pi_2(M^{2n})=0$.
 The class $[\omega]$ is nontrivial in $H^2(M^{2n})$,
i.e. $[\omega]\not=0$, since
cohomology class of the volume form $\omega^n$ is $n$ times the
class $[\omega]$ and it is nontrivial since $V\not=0$.
 If we let $\pi_1(M^{2n})=0$, we have the isomorphism
$\pi_2(M^{2n})\simeq H_2(M^{2n},Z)$ according to the Gurevich
theorem. Therefore, according to the de Rham theorem the
condition (\ref{intomega}) leads to $[\omega]=0$, that contradicts
to $V\not=0$.

 In other words, necessary condition for unbroken time
reparametrization invariance of the model is that there should
be non-contractible loops in $M^{2n}$ for which, particularly,
the Wilson loop integrals build up a representation.
 In Sec.~\ref{corfun}, we show how one can construct
a term leading to broken time reparametrization invariance
for compact $M^{2n}$ with nontrivial $\pi_1(M^{2n})$.

 Below, we turn to some basic notions of the supersymmetric model
relevant to subsequent consideration.

 The most basic elements of the given $d=1$, $N=2$ theory are
analogues of the chiral and anti-chiral rings of the (topologically
twisted) two-dimensional $N=2$ models\cite{Cecotti-91}.
 Since we are originally interested in the topologically twisted
$N=2$ model only the ground states are kept, so it is simple to
make identification of these with the operators.
 Namely, the chiral operators, $\phi_i$, are defined as the ones
satisfying $[Q_{\beta},\phi_i]=0$, and the anti-chiral operators,
$\bar\phi_i$, satisfy $[\bar Q_{\beta},\bar\phi_i]=0$.
 They are irreducible representations of the supersymmetry algebra.
 Then, we can make a correspondence between the Ramond ground
states defined by (\ref{Qbetarho}) and chiral fields by choosing
a canonical ground state $|0\rangle$, with the identification
$\phi_i|0\rangle =|i\rangle +Q_{\beta}|\lambda\rangle$.
 Similarly, there is a natural isomorphism between the anti-chiral
fields and the adjoint states $|\bar i \rangle$.
 In terms of the Landau-Ginzburg formulation, the chiral ring
consists of the polynomials modulo the relation $dW=0$, which
defines critical points of the Hamiltonian vector field.

 The inner product on the space of the ground states corresponding
to the fields $\phi_i$ and $\phi_{\bar j}$ is
$g_{i\bar j} = \langle\bar j |i \rangle$, and geometrically plays
the role of a metric in the associated Hilbert space while
$\langle i |j \rangle$ gives rise to the topological metric,
$\eta_{ij} = \langle \phi_i \phi_j \rangle_{top}$, which will be
discussed in Sec.~\ref{correlation}, in the context of correlation
functions of BRST observables. The real structure matrix,
$M = g\eta^{-1}$, relates the ground states with its adjoints,
$\langle \bar k| = \langle j| M^{j}_{\bar k}$.

\subsection{DEFORMATIONS AND PERTURBATIONS}\label{Dfrm}

 Within the Batalin-Vilkovisky formalism
(see for a review Ref.\cite{Henneaux-85}),
recent general analysis by Anselmi\cite{Anselmi-93} of the
predictivity and renormalizability of (reducible and irreducible)
topological field theories which are known to be entirely determined
by the gauge fixing (the classical action is either zero or
topological invariant), shows that any topological field theory is
predictive. The central point for that theories is thus
the gauge fixing; for example, two gauge fixings which can not be
continuously deformed one into the other give rise to
inequivalent theories.

 In the case under study, the gauge fixing condition is Hamilton's
equation whose deformations should be studied in order to garantee
correctness of the definition of observables of the theory.
 Also, it would be interesting to study symmetry preserving
perturbations of the action (\ref{I}) in order to find metric of the
supersymmetric ground state space
(see discussion in Sec.~\ref{discussion}) and possible deformations
of the cohomology ring ${\cal R}$ of observables.
 However, there seems to be no nontrivial deformations of
the cohomology ring since at least there is neither "quantum" nor
conventional instanton corrections to the linear $d=1$ theory.

 In general, the supersymmetry preserving F-term perturbation of
the action can be written using the chiral and anti-chiral fields,
\begin{equation}
                                                        \label{pert}
\delta I = \sum_k  \delta      t^k \int dt
           \{\bar Q_{\beta},[\bar Q_{\beta},\phi_k]\}
         + \sum_k  \delta \bar t^k \int dt
           \{     Q_{\beta},[     Q_{\beta},\bar\phi_k]\},
\end{equation}
where $\delta t^k$ and $\delta\bar t^k$ are coupling parameters.
 Since we are interested in Hamiltonian systems, we leave the
form of Hamilton's equation unchanged, and it is suffice for our
purpose to look at the deformations of (i) Hamiltonian and
(ii) symplectic tensor entering Hamilton's equation,
to identify which type of them preserves the symmetries of the
theory.

(i) Let us consider deformation of the Hamiltonian,
\begin{equation}
                                                        \label{defH}
H(a) \rightarrow H(a) - \delta t_{P}P(a),
\end{equation}
where $P(a)$ is a local polynomial, and $\delta t_{P}$ is a coupling
constant parametrizing the deformation.
 This leads immediately to the following perturbation of the action
(\ref{I}):
\begin{equation}
                                                        \label{intL}
\int dt{\cal L} \rightarrow  \int dt {\cal L} +
                 \delta t_{P}\int dt {\cal O}_P,
\end{equation}
where
\begin{equation}
                                                        \label{OP}
{\cal O}_P =
 iq_i \omega^{ij}\partial_j P(a)
 - \bar c_i \omega^{ij}\partial_k\partial_j P(a)c^k.
\end{equation}
 Direct calculations show that ${\cal O}_P$
is BRST and anti-BRST closed, $s{\cal O}_P = \bar s{\cal O}_P = 0$;
see (\ref{QOP}) below, with $\beta=0$.

 The matter of an immediate interest is whether the possible
deformations of the superpotential are in one-to-one correspondence
with the possible topological perturbations of the theory,
as it is, for example, in the Landau-Ginzburg formulation of
$d=2$, $N=2$ superconformal field theories.
 Non-trivial topological perturbation may have place only if
the deformation term is not a BRST exact cocycle.
 It can be readily checked that, in our case,
\begin{equation}
                                                        \label{OPs}
{\cal O}_P = s \{\bar c_i\omega^{ij}\partial_j P(a)\},
\end{equation}
so that there are no {\sl nontrivial} topological perturbations
coming from the deformation of the superpotential.
 In other words, nothing is changed in the topological sector
when one deforms the superpotential by local polynomial.
 This result confirms our remark concerning the homotopical
character of the topological twist (\ref{Qbeta}).

 However, supersymmetry appears to be sensitive to the deformation.
 We now examine the condition for the deformation ${\cal O}_P$
to be supersymmetry preserving.
 Using the definitions (\ref{Qbeta}) and (\ref{OP}) we find directly
\begin{equation}
                                                        \label{QOP}
[Q_{\beta}, {\cal O}_P] =
-i\beta c^k\partial_k(\omega^{ij}\partial_i H \partial_j P),
\quad
[\bar Q_{\beta}, {\cal O}_P] =
i\beta \omega^{mn}\bar c_m\partial_n
(\omega^{ij}\partial_i H \partial_j P).
\end{equation}
 Sufficient condition to both the commutators in (\ref{QOP}) vanish
is that the Poisson bracket
\begin{equation}
                                                        \label{pb}
(\partial_i H)\omega^{ij}(\partial_j P)
\equiv \{H,P\}_{\omega} = \mbox{const},
\end{equation}
or, more precisely, is equal to a locally constant function.
 For the case of linearly connected $M^{2n}$, equation (\ref{pb})
is necessary and sufficient condition for the Hamiltonian flows
defined by the
functions $P$ and $H$ to commute, with $P$ viewed as another
Hamiltonian\footnote{For the phase space with nontrivial $\pi_1$ one
should use here local Hamiltonian flows.}.
 We note that, in general, when one knows a Hamiltonian
flow commuting with the flow under study it is possible to construct
an integral of motion\cite{Abraham-Arnold}.
 Hence, the equations (\ref{QOP}) represent a link between
the supersymmetry and integrability.

 When examining the formal evolution of $P$, we see that $P$ linearly
changes with time, $dP/dt=$ const.
 For compact connected $M^{2n}$ polynomial $P(a)$ is bounded so
that the constant must be zero, {\sl i.e.} $P$ is an integral of
motion, $\{H,P\}_{\omega} = 0$.
 It has been argued\cite{GR-90,GRT-91c} that the existence of an
additional {\sl nontrivial} integral of motion leads to
{\sl broken} supersymmetry.
 This argument is based on analysis made on the form of
the supersymmetric ground state (\ref{Gibbs}).
 In view of this, polynomial $P$ should be a trivial integral of
motion to preserve the supersymmetry.

 For noncompact $M^{2n}$, the polynomial $P$ is not necessarily
bounded so that we are left with the general condition (\ref{pb}).
 The same is true for $M^{2n}$ with nontrivial $\pi_1$ and
also when $P$ is an analytic function.
 However, in general, if (polynomial or analytic function)
$P$ is in involution with $H$ it must be a {\sl trivial} first
integral.

 From the above analysis we conclude that the symmetries do
not fix the Lagrangian uniquely, with nontrivial supersymmetry
preserving perturbation term (\ref{OP}), where $P$ satisfies
(\ref{pb}) with {\sl non-zero} constant, can be added to the action.
 However, in the case of compact connected $M^{2n}$ there are no
nontrivial supersymmetry preserving perturbations supplied by the
deformation with polynomial.

 (iia) Let us turn to considering of deformation of the constant
symplectic tensor. Under an infinitesimal change
$\omega_{ij} \to \omega_{ij}+\epsilon_{ij}$,
one sees from (\ref{Qbeta}) that $Q_{\beta}$ is invariant whereas
$\bar Q_{\beta}$ changes by
\begin{equation}
                                                     \label{QbetaX}
\delta\bar Q_{\beta}
= [Q_{\beta}, \bar K_{\epsilon}]
= \bar c_i\epsilon^{ij}(iq_j+\beta\partial_j H),
\end{equation}
with
\begin{equation}
                                                     \label{Komega}
\bar K_{\epsilon}
=\frac{1}{2}\epsilon^{ij}\bar c_{i}\bar c_{j},
\end{equation}
where
$\epsilon_{ij}\epsilon^{jk}=\delta^k_i$.
 Using (\ref{susy}) and (\ref{QbetaX}) one finds that the
Hamilton function changes by
\begin{equation}
                                                     \label{QQK}
\delta{\cal H}
=\frac{1}{2i\beta}
\{Q_{\beta},[Q_{\beta}, \bar K_{\epsilon}]\}
=q_i\epsilon^{ij}\partial_j H
+i\bar c_i\partial_k(\epsilon^{ik}\partial_j H) c^j.
\end{equation}
 In order to preserve the $N=2$ supersymmetry algebra,
$\bar K_{\epsilon}$ must commute with $\bar Q_{\beta}$,
\begin{equation}
                                                   \label{QbetaK}
[\bar Q_{\beta}, \bar K_{\epsilon}]
= \frac{1}{2}\omega^{kl}\partial_l\epsilon^{ij}\bar c_i\bar c_j\bar c_k,
\end{equation}
where we have used $q_l=-i\partial_l$.
The {\sl r.h.s.} of (\ref{QbetaK}) vanishes if and only if
$\epsilon_{ij}$ are antisymmetric and constant so that
it appears to be the case of a variation of the symplectic
structure. As a consequence, this variation preserves also
the BRST and anti-BRST symmetries.
 We notice that the form of Eqs.(\ref{Komega}) and (\ref{QQK})
is very suggestive to represent the Hamilton function in the form
\begin{equation}
                                                     \label{HQQK}
{\cal H} = \frac{1}{2i\beta}\{Q_{\beta},[Q_{\beta}, \bar K]\}.
\end{equation}
 This representation stems naturally from combination of
supersymmetry algebra (\ref{susy}) and ISp(2) algebra (\ref{ISP2}),
and thus is specific to the model.

 (iib) When one attempts change by {\sl non-constant}
tensor, $\epsilon_{ij}=\epsilon_{ij}(a)$, the previous arguments
break down because the second equality in (\ref{QbetaX}) does not
hold, and, even more, the anti-BRST operator receives non-nilpotent
contribution.
 This case, however, is important since, as it was mentioned above,
constant symplectic tensor can not be in general globally defined
on a symplectic manifold. For instance, on a compact one, for which
case one uses a covering by local charts with constant
$\omega_{ij}$ owing to Darboux theorem telling us that in some
neighborhood of any point one can find local coordinates such that
$\omega=dx^{\alpha}\wedge dp_{\alpha}$.
 Also, a reasonable expectation is that this might yield
a mechanism for supersymmetry breaking.

 So, we are led to consideration of the model with a coordinate
dependent symplectic structure, $\omega_{ij}=\omega_{ij}(a)$,
so called Birkhoffian mechanics, in which one does not use the
Darboux coordinates and attempts to treat symplectic structure
in a full generality.
 Analysis made on this generalized
model\cite{GRT-92,Aringazin-93-HJ,Aringazin-93-PL} has shown that
with the following modification of the anti-BRST operator,
\begin{equation}
                                                       \label{barBQ}
\bar Q = i\bar c_{i}\omega^{ij}(a)q_{j} -
\frac{1}{2}\partial_{k}\omega^{ij}(a) c^{k}\bar c_{i}\bar c_{j},
\end{equation}
obtained by virtue of $[\bar K,Q]=\bar Q$, the ISp(2) algebra
(\ref{ISP2}) is regained, with all the basic results of the constant
symplectic structure case being reproduced.

 The BRST approach to this generalized model can be readily
developed in the same fashion as it for the case of Darboux
coordinates.
 Besides slight modifications, which do not influence the algebraic
structure of the original model, we encounter the following
remarkable difference.
 According to the modification (\ref{barBQ}) the supercharge
in (\ref{Qbeta}) can be brought to the form
\begin{equation}
\bar Q_{\beta} = \bar c_{i}D^{i} -                \label{barBQHD}
\frac{1}{2}f^{kl}_{\ \ m}c^{m}\bar c_{k}\bar c_{l},
\end{equation}
where
\begin{equation}
                                                     \label{D+}
D^{i} = \omega^{ij}(a)(\partial_j + \beta\partial_j H)
\end{equation}
and
\begin{equation}
                                              \label{f}
f^{ijk} =  \omega^{im}(a)\partial_{m}\omega^{jk}(a)
          -\omega^{jm}(a)\partial_{m}\omega^{ik}(a).
\end{equation}
Our observation is that, in the BRST approach to gauge field
theories, the operators placed similarly as $D^{i}$ in
(\ref{barBQHD}) play the role of generators of Lie group
characterizing gauge symmetry of the theory, and $f^{kl}_{\ \ m}$
are the structure constants.
 It is easy to check that $D^{i}$ fulfills the commutation rule
\begin{equation}
[ D^i, D^j ] = f^{ijk}D_k,                   \label{comD}
\end{equation}
and, owing to Jacobi identity of the Poisson bracket algebra,
$\omega^{im}\partial_{m}\omega^{jk} +
\omega^{km}\partial_{m}\omega^{ij} +
\omega^{jm}\partial_{m}\omega^{ki} = 0$,
{\sl i.e.} $d\omega(a)=0$,
$f^{ijk}$ satisfies
\begin{equation}
f^{ijk} + f^{kij} + f^{jki} = 0,                    \label{fff}
\end{equation}
so that the operators $D^{i}$ constitute a Lie
algebra\footnote{We assume that $f^{ijk}$'s are local constants.}.
 Note that the algebra defined by (\ref{comD}) is {\sl degenerate} in
Darboux coordinates of Hamiltonian mechanics, in which case
we have identically $f^{ijk} = 0$.

 Now, with the infinitesimal change,
$\omega_{ij}(a) \to \omega_{ij}(a)+\epsilon_{ij}(a)$,
the first equalities in Eqs. (\ref{QbetaX}) and (\ref{QQK}),
where ${\cal H}$ and $\bar Q_{\beta}$ are defined by
(\ref{HQQK}) and (\ref{barBQ}) respectively, are still valid
while (\ref{QbetaK}) becomes
\begin{equation}
                                                     \label{QbetaK2}
[\bar Q_{\beta}, \bar K_{\epsilon}]
= \frac{1}{2}\{\omega^{kl}(a)\partial_l\epsilon^{ij}(a)
              +\epsilon^{kl}(a)\partial_l\omega^{ij}(a)\}
\bar c_i\bar c_j\bar c_k,
\end{equation}
with the result is that, again, closeness of the two-form
$\epsilon$ is sufficient for the {\sl r.h.s.} of (\ref{QbetaK2})
to be zero, and thus the $N=2$ supersymmetry to be preserved.

 It is highly remarkable, however, that the above condition is
equivalent to the one that the following Schouten
bracket\cite{Schouten-40} is zero,
\begin{equation}
                                                   \label{Schouten}
[[\omega,\epsilon ]]^{kij} \equiv
   \sum\limits_{(kij)}
  (\omega^{kl}\partial_l\epsilon^{ij}
  +\epsilon^{kl}\partial_l\omega^{ij}) = 0,
\end{equation}
which is necessary and sufficient condition for $\omega$ and
$\epsilon$ to be a Poisson pair\cite{Karasev-91}, {\sl i.e.} for
$k_1\omega + k_2\epsilon$ to be a two-parameter family of tensors
defining a Poisson bracket on $M^{2n}$.

 So, the general result both for (iia) and (iib) is that the $N=2$
supersymmetry is preserved under the deformation when Schouten
bracket between $\omega$ and $\epsilon$ is zero.

 The following comments are in order.

 (i) We see that the supersymmetry imposes nontrivial condition
(\ref{Schouten}) for deformation $\epsilon$ of the original Poisson
bracket. Indeed, locally or globally, there may be both trivial and
nontrivial deformations. Clearly, the class of global nontrivial
deformations is related to topology of $M^{2n}$ and, thus, is most
interesting to investigate.

 (ii) Also, one can study the anomalies,
\begin{equation}
                                                   \label{susybreak}
[[\omega,\epsilon ]]=\Gamma,
\end{equation}
where antisymmetric rank-three tensor $\Gamma$ measures supersymmetry
breaking. We emphasize that, in general, this provides very
attractive mechanism for supersymmetry breaking.

 (iii) In some cases such anomalies may come naturally. Namely, it is
known that some of nonlinear Poisson brackets describing dynamics of
physical systems can be made linear by appropriate deforming original
$\omega$. Generally, it looks like one attempts a deformation inside
the usual Poisson bracket so that we have not to extend our study for
nonlinear Poisson bracket case. Note that such deformations are not
trivial, at least locally. Particularly, in some cases they are
parametrized by a set of parameters, and, as the supersymmetry is
related to the motion regimes, one can use criterion $\Gamma=0$ to
find critical values of the parameters distinguishing between the
regular and nonregular regimes. We will not discuss here specific
examples which can be made elsewhere.

\subsection{EQUIVARIANT EXTERIOR DERIVATIVE}\label{Eqvr}

 In this Section, we briefly present construction of the
model under study by the use of a generalized Mathai-Quillen
formalism\cite{Niemi-94}.

 Clear geometrical meaning of the model suggests that its
constructing can be refined using an {\sl equivariant} exterior
derivative.
 The generalized Mathai-Quillen formalism appeared to be
useful\cite{Niemi-94} in analyzing supersymmetry
properties of models describing classical dynamical systems,
in an exterior calculus framework.
 Particularly, this technique can be used to construct the
supersymmetric models for Hamiltonian systems which are not
necessarily of cohomological type.
 Also, it provides a relevant basis to attempt breaking of
supersymmetry, which appeared to be concerned to motion regimes
discussed in Sec.~\ref{Dfrm}. However, we will not try to
use it for this purpose here, restricting our investigation on
setting up the formulation.

 Since the fields $a^i$ and $\bar c_i$ can be viewed as
local coordinates of the cotangent bundle $T^*M^{2n}$
the corresponding basic one-forms can be identified
with $c^i$ and $q_i$ respectively; see Table~\ref{table}.
 The nilpotent exterior derivative on the exterior
algebra in the space of mappings from circle $S^1$ to
$T^*M^{2n}$ is thus given by
\begin{equation}
                                                   \label{d}
d = \int dt
\Bigl(
  c^i \frac{\partial}{\partial     a^i}
+ q_i \frac{\partial}{\partial\bar c_i}
\Bigr).
\end{equation}
 Comparing (\ref{d}) with (\ref{s}) we see that the exterior
derivative $d$ and the BRST operator $\int dt\ s$ are
equivalent to each other (herebelow, we omit $i$ factors for
simplicity).

 By introducing the interior multiplication operator
along the vector field $v$,
\begin{equation}
                                                   \label{v}
v=(\partial_t a^i-h^i,\, \partial_t \bar c_k+\bar c_j\partial_kh^j),
\end{equation}
namely,
\begin{equation}
                                                   \label{iv}
i_v
=\int dt
\Bigl(
 (\partial_t a^i-h^i) \tau_i
+(\partial_t \bar c_i+\bar c_j\partial_kh^j) \pi^i
\Bigr).
\end{equation}
where $\tau_i$ and $\pi^i$ form the basis of contractions
dual to $c^i$ and $q_i$ respectively, {\sl i.e.}
\begin{equation}
                                                   \label{tauc}
\tau_i c^k =\delta^k_i\delta(t-t'),
\quad
\pi^i q_k =\delta^i_k\delta(t-t'),
\end{equation}
we define the following equivariant exterior derivative
\begin{equation}
                                                   \label{Qv}
Q_v \equiv d + i_v
=\int dt
\Bigl(
c^i \frac{\partial}{\partial     a^i}
+q_i \frac{\partial}{\partial\bar c_i}
+(\partial_t a^i-h^i) \tau_i
+(\partial_t \bar c_i+\bar c_j\partial_kh^j) \pi^i
\Bigr).
\end{equation}
 Note that the second component of the vector field $v$ is Jacobi
variation of the first one.
 The corresponding Lie derivative is given by
\begin{equation}
                                                   \label{elldi}
\ell = Q^2_v,
\end{equation}
so that according to (\ref{Qv})
\begin{equation}
                                                   \label{ellv}
\ell
=\int dt
\Bigl(
 \partial_t      a^i \frac{\partial}{\partial     a^i}
+\partial_t \bar c_i \frac{\partial}{\partial\bar c_i}
+\partial_t      c^i \tau_i
+\partial_t      q_i \pi^i
+\ell_h
\Bigr),
\end{equation}
and hence $\ell=\int dt\ (\partial_t+\ell_h)$,
where $\ell_h$ is a Lie derivative along $h^i$, and it is obvious
that
\begin{equation}
                                                   \label{ellQ}
[\ell, Q_v] = 0.
\end{equation}
 Eqs.(\ref{elldi}) and (\ref{ellQ}) constitute a superalgebra.
 We see that (\ref{ellv}) is the operator corresponding to the
Liouville equation of classical mechanics.
 Action of the equivariant exterior derivative (\ref{Qv})
on contraction of the basic one-forms,
\begin{equation}
                                                   \label{B''}
B'' = c^iq_i,
\end{equation}
yields the action,
\begin{equation}
                                                   \label{Iv}
I = Q_v B''
  = \int dt
\Bigl(
  q_i(\partial_t a^i-h^i)
+ \bar c_i\partial_t c^i
+ \bar c_i\partial_j h^i c^j
- \partial_t(\bar c_i c^i)
\Bigr),
\end{equation}
where (\ref{tauc}) has been used.
 It is equivalent to the original action with the Lagrangian
(\ref{L'}).
 Zeroth of the $\tau_i$ component of the vector field in
(\ref{Qv}) are solutions to $\partial_t a^i = h^i$  so that
the action (\ref{Iv}) describes these field configurations.
 Comparing the derivation of (\ref{Iv}) with the one of the BRST
scheme, we see that the trick provided by this technique is that
the gauge fixing condition, $\partial_t a^i-h^i=0$, is encoded
in the equivariant exterior derivative $Q_v$ rather than it is
described by the "gauge fermion" $B''$ and thus, in contrast to
$s$, the operator $Q_v$ itself carries information on the
dynamical system.

\section{BRST OBSERVABLES AND CORRELATION FUNCTIONS}
\label{correlation}

\subsection{BRST OBSERVABLES}\label{brstobs}

 In general, observables of interest are of the form
\begin{equation}
                                                          \label{O}
{\cal O}_{A} = A_{i_{1}\cdots i_{p}}(a)c^{i_{1}}\cdots c^{i_{p}},
\end{equation}
which are $p$-forms on $M^{2n}$, $A \in \Lambda^{p}$.

In general, the space of $p$-forms on a manifold equipped by Poisson
bracket has a structure of Lie superalgebra in respect to the
following Karasev bracket (supercommutator) between the
forms\cite{Karasev-91}:
\begin{equation}
                                                    \label{Karasev}
[A,B]_K = d\omega(A,B) +\omega(dA,B) +(-1)^{deg(A)+1}\omega(A,dB),
\end{equation}
where
\begin{equation}
                                                    \label{omegaAB}
\omega(A,B)_{i_1\cdots i_{m+n-2}} =
\sum\limits_{(i'_1,\cdots,i'_{m+n-2})}
(-1)^{e(i'_1,\cdots,i'_{m+n-2})}
A_{ri'_1\cdots i'_{m-1}}\omega^{rs}B_{si'_m\cdots i'_{m+n-2}},
\end{equation}
the sum is over all cyclic permutations, and $e(\cdots)$ denotes
index of permutation.
 This is an algebra of observables in our case\footnote{Note that
Karasev bracket (\ref{Karasev}) for forms corresponds to Schouten
bracket for associated antisymmetric tensors.}.

 However, one can easily find that for the BRST invariant observables
$\{ Q,{\cal O}_A\} = 0$ if and only if $A$ is closed since
$\{ Q,{\cal O}_A\} = {\cal O}_{dA}$.
 Consequently, the BRST observables correspond to the de Rham
cohomology and form a classical cohomology ring ${\cal R}$ of
$M^{2n}$ which corresponds to the ring of chiral operators $\phi_i$.
 So, for the BRST observables on a symplectic manifold the Lie
superalgebra defined by (\ref{Karasev}) is {\sl trivial} since
for closed $\omega$, $A$, and $B$ we have identically $[A,B]_K = 0$.

 The BRST observables are related immediately to the BRST invariant
states, via construction analogous to that of relating the chiral
fields to the supersymmetric ground states made in Sec.~\ref{LG}.
 The distribution forms are identified with differential forms as
\begin{equation}
                                                          \label{A}
 A_{i_{1}\cdots i_{p}}(a)c^{i_{1}}\cdots c^{i_{p}}|0\rangle
                         \leftrightarrow
 A_{i_{1}\cdots i_{p}}(a)da^{i_{1}}\wedge \cdots \wedge da^{i_{p}}.
\end{equation}
 In terms of the vacuum distribution forms, the isomorphism between
the chiral fields and states in the Ramond sector becomes more
explicit.
 Namely, the Hilbert space of the model consists of all square
summable $p$-forms, $|A_i\rangle = |i\rangle$, with coefficients
taking value in some linear bundle $E$ on which the operators
$\phi_i$ corresponding to the cohomology classes act by wedge
product.

 We note that since the flow equation is real the complex conjugate
of the vacuum distribution form $A_i$ should be again a vacuum
distribution form, and thus can be expressed as a linear
combination of the vacuum distribution forms.

\subsection{HOMOTOPY CLASSES OF THE FIELDS}\label{hmtp}

 The basic field $a^i(t)$ is characterized by homotopy classes
of the map $M^1 \rightarrow M^{2n}$.
 Clearly, these are in general classes of the map
$S^1 \to M^{2n}$, that is the classes of conjugated elements of
the fundamental homotopy group.
 These classes can be weighted with different phases and
controlled by some coupling parameter.
 Namely, let us consider the space ${\cal E}(a_0,a_1)$ of the fields
$a^i(t)$ coinciding with $a^i_0$ at $t=t_0$ and with $a^i_1$
at $t=t_1$. The functional integral (\ref{Z}) is performed
over histories ${\cal E}(a_0,a_1)$, and can be presented as
\begin{equation}
                                                     \label{Zalpha}
Z = \sum\limits_{\alpha}e^{ig_{\alpha}}
              \int\limits_{{\cal E}(a_0,a_1)}
               D[X]\exp\ iI,
\end{equation}
where $g_{\alpha}$ is the phase and $\alpha$ runs over components
of ${\cal E}(a_0,a_1)$. The case $a_0\not= a_1$ can be reduced
to the case $a_0=a_1$ since ${\cal E}(a_0,a_1)$ is either
homotopically empty, or homotopically equivalent to
${\cal E}={\cal E}(a_0,a_0)$.
 Consequently, the components of ${\cal E}(a_0,a_1)$ are
in one-to-one correspondence with elements of $\pi_1(M^{2n},a_0)$.
Furthermore, since the spaces of fields $a^i(t)$ at different
fixed $t$ are trivially equivalent to each other, the groups
$\pi_1(M^{2n},a_0)$ at different $a^i_0$ are isomorphic
to each other. So, we are led to consider closed paths,
$a^i(t_0)=a^i(t_1)$, which are elements of the loop space
${\cal E}$ or,
equivalently, fields on a circle, $t\in S^1$, with the index
$\alpha$ in (\ref{Zalpha}) running over
$\pi_1(M^{2n})$\footnote{For a complete definition, appropriate
boundary condition for ghosts should be specified as well.}.
 Physically, as the energy of the system is finite,
different homotopical classes of the fields can be thought of
as they are separated by infinitely high energy barriers.

 The fields are thus characterized by appropriate representations
$\sigma$ of the group $\pi_1(M^{2n})$, which we assume to be
nontrivial, partially for the reason mentioned in Sec.~\ref{LG}.
 Therefore, the coefficients of the cohomology ring ${\cal R}$ take
values in the linear bundles $E_{\sigma}$ associated to the
representations $\sigma$.

 Thus, we should study specifically periodic orbits in
$M^{2n}$ characterized by period $T = |t_1-t_0|$.
 The $T$-periodic solutions to Hamilton's equation are
elements of the loop space of Hamiltonian system which is a subject
of recent studies\cite{Niemi-94,Hofer-94,Floer-86} on infinite
dimensional version of Morse theory.
 We remark that periodic orbits are presumably dense in phase
space and at finite time scale may mimic typical dynamics
arbitrary well. Moreover, the families of periodic orbits
have the unique property that they continue smoothly across
the fractal boundary between the regular region and the
chaotic region, with stable and unstable character in these
regions respectively, being thus the only unifying agents
between these two disparate regions.

 When necessary one can replace circle by the real line by taking
the limit $T \to \infty$.
 This procedure is useful from a general point of view, and,
particularly, it allows one to extract\cite{GR-92} Lyapunov
exponents, positive values of which are wellknown to be a strong
indication of chaos in Hamiltonian systems, from correlation
functions.

\subsection{CORRELATION FUNCTIONS}\label{corfun}

 Let us now turn to consideration of the correlation functions
of the BRST invariant observables (\ref{O}).
 If $N$ is a closed submanifold of (compact) $M^{2n}$ representing
some homology class of codimension $m$ ($2n$-$m$ cycle), then,
by Poincare duality, we have $m$-dimensional cohomology
class $A$ ($m$-cocycle), which can be taken to have
delta function support on $N$\cite{Bott-82}.
 Thus, any closed form $A$ is cohomologious to a linear
combination of the Poincare duals of appropriate $N$'s.
The general correlation function is then of the form,
\begin{equation}
                                                   \label{corO}
 \langle {\cal O}_{A_1}(t_1) \cdots  {\cal O}_{A_m}(t_m)
 \rangle,
\end{equation}
where $A_k$ are the Poincare duals of the $N$'s.
 Our aim is to find the contribution to this correlation
function on $S^1$ coming from a given homotopy class of
the map $S^1\rightarrow M^{2n}$.
 The conventional techniques with the moduli space
${\cal M}$\cite{Witten-88} consisting of the fields $a^i(t)$
of the above topological type can be used here owing to the BRST
symmetry. Namely, the non-vanishing contribution to (\ref{corO})
can only come from the intersection of the submanifolds
$L_k \in {\cal M}$ consisting of $a$'s such that $a^i(t)\in N_k$,
and we obtain familiar formula\cite{Witten-91},
\begin{equation}
                                             \label{intersection}
\langle {\cal O}_{A_1}(t_1) \cdots
        {\cal O}_{A_m}(t_m)\rangle_{S^1}
    =  \#\Bigl(\sum\limits^m\cap L_k\Bigr),
\end{equation}
relating the correlation function to the number of intersections.
 As it was expected, the correlation functions do not depend on
time but only on the indices of the BRST observables. These results
are typical for all topological field theories.
 Now we turn to some specific results following from this
consideration.

{\sl Homotopically trivial sector.}

 This sector is characterized by $a$'s which are homotopically
constant maps, $[a^{i}(t)] = [a^{i}(t_{0})]$, and, therefore,
the correlation function (\ref{intersection}) can be presented as
\begin{equation}
                                              \label{intersection2}
\int A_1\wedge \cdots \wedge A_m.
\end{equation}
 The standard Poincare integral invariants, $K^p$, identified
in Sec.~\ref{Chmlg} as fundamental BRST observables correspond to
this homotopically trivial sector since they have been originally
formulated in the Darboux coordinates, in which $\omega_{ij}$
are constant coefficients.
 Also, two-point correlation function can be used to define the
topological metric,
$\eta_{ij} \equiv \langle A_i|A_j\rangle = \int A_{i} \wedge A_{j}$,
which is just the intersection form in the cohomology
(cf. Ref.\cite{Cecotti-92}).
 Particularly, it is easy to check that in canonical basis of the
forms, the action of the real structure matrix $M$ reads
$*A^*_j = g_{i\bar j}A_i$, and the topological metric is
$\eta_{ij}=\delta_{ij}$.

Note that circle $S^1$ is mapped by $a^i(t)$ to some
one-dimensional cycle, $C(a)\subset M^{2n}$, associated to the field.
Using this cycle and a closed two-form, $\psi$, one can construct
a multivalued term which can be added to the action of the model.
Namely,
\begin{equation}
                                                        \label{Ipsi}
I_{\psi}=2\pi\int\limits_{\gamma(a)}\psi,
\end{equation}
where $\gamma(a)$ is an arbitrary two-dimensional surface, for which
$C(a)$ is its border, $\partial\gamma(a)=C(a)$.
 In general, such a surface may not exist
since the cycle $C(a)$ may not be homological to zero.
So, we restrict our consideration to the case when
such a surface exists. This may be done either by imposing
topological restriction $\pi_1(M^{2n})=0$ on the phase space
that we still avoid to accept,
or by considering homotopically trivial class of fields $a^i(t)$,
for which $C(a)$ is homologically zero.

 Clearly, the value of $I_{\psi}$ may depend on the choice of
$\gamma(a)$. However, when $\psi$ is an {\sl integer valued}
two-form, {\sl i.e.}
\begin{equation}
                                                     \label{intpsi}
\oint_{\gamma}\psi = k,
\qquad \gamma \subset H_2(M^{2n},Z),
\end{equation}
where $\gamma$ is an arbitrary two-dimensional cycle and
$k$ is an integer number,
then any two choices of $\gamma(a)$ in (\ref{Ipsi}) differ by
$2\pi k$ so that $\exp\ iI_{\psi}$ is univalued.

   Thus, from the point of view of a functional formulation of
field theory, for integer valued closed two-form $\psi$ the
term (\ref{Ipsi}) is well defined.
Such a term can be added to the original action, and may play
important role when analyzing symmetries of the model.
Particularly, we expect that with an appropriate choice of
the cocycle $\psi$ it can be used to break some of the symmetries.
 For example, in the case of low-energy limit of QCD
inclusion of such a topological term provides breaking of
excessive symmetry of Goldstone fields to meet experimental data.

 One of the candidates for $\psi$ is properly symplectic two-form
$\omega$. One can show that in this case time reparametrization
invariance of the model for compact $M^{2n}$ is necessarily broken.
Indeed, $I_{\psi}$ with $\psi=\omega$ breaks the time
reparametrization invariance unless the condition
$\oint_{\gamma}\omega = 0$ is satisfied because $\omega$ should
obey both of (\ref{intpsi}) and (\ref{intomega}).
The latter condition entails that the cohomological class of
$\omega$ is zero. However, for compact $M^{2n}$ this class is
necessarily nonzero.\\

{\sl Homotopically nontrivial sectors.}

 Existence of homotopically nontrivial Poincare invariants, $K^p$,
follows from the fact that,
globally, $\omega_{ij}(a)$ may not be chosen constant, and
there are nontrivial homotopy classes of $a^i$.
 Nontrivial character of these invariants comes from the fact
that cohomology class of $\omega$ on compact $M^{2n}$ is
necessarily nontrivial.

 Let us turn to a particular kind of observables
intertwining the BRST and anti-BRST ones.
 The BRST observable (\ref{O}) can be naturally understood
as $(p,0)$-form corresponding to the general $(p,q)$-form,
\begin{equation}
                                                       \label{U}
{\cal U}_A =
A_{i_1\cdots i_p}^{j_1\cdots j_q}(a) c^{i_1}\cdots c^{i_p}
\bar c_{j_1} \cdots \bar c_{j_q}|0\rangle
          \leftrightarrow
A_{i_1\cdots i_p}^{j_1\cdots j_q}(a)
da^{i_1}\wedge\cdots\wedge da^{i_p}
\partial_{j_1}\wedge\cdots\wedge\partial_{j_q},
\end{equation}
which can be viewed as a function on $M^{4n|4n}$, where the
anti-ghosts represent the anti-BRST sector.
 Accordingly, we associate the $(0,q)$-forms to the anti-BRST
observables, which can be treated in the same manner as the BRST
ones.

 A particular kind of the observable (\ref{U}) intertwining the BRST
and anti-BRST sectors has been studied recently by Gozzi and
Reuter\cite{GR-92},
\begin{equation}
                                                     \label{OGozzi}
{\cal U}_A = \delta(a(t_0)-a_0)
     c^{i_1}(t)   \cdots      c^{i_p}(t)
\bar c_{i_1}(t_0) \cdots \bar c_{i_p}(t_0)|0\rangle.
\end{equation}
 After normal ordering, the observable (\ref{OGozzi}) can be thought
of as the operator creating $p$ ghosts from the Fock vacuum ($p$-volume
form in $TM^{2n}$) at some time $t_{0}$ and point $a_{0} \in M^{2n}$,
and then annihilating them at some later time $t$.
 Certainly, we should arrange also time-ordering to define this
operator correctly. However, we have not to specify the time
in the associated correlation function since we dealing with
$t \in S^1$.
 Indeed, the correlation function for (\ref{OGozzi}),
\begin{equation}
                                                     \label{Gamma}
 \langle {\cal U}_{A}\rangle_{S^1}
      = \langle \bar A(t) A(t_0)\rangle_{top}
 \equiv \Gamma_p (T, a_0),
\end{equation}
does not depend on specific time, and $\Gamma_p$ depends only on
the period $T$.

 A nice result\cite{GR-92} is that the higher order
($p \geq 1$) largest Lyapunov exponents can be extracted from
this correlation function, namely,
\begin{equation}
                                                   \label{Lyapunov}
 l_{p}(a_{0}) = \sum\limits^{p}_{m=1}
                \lim_{T \to \infty}
                \sup \frac{1}{T}ln\, \Gamma_{m}(T, a_{0}).
\end{equation}

 The $p$-form sector of the partition function (\ref{Z}) with
appropriate periodic boundary conditions and the fields defined on
circle,
\begin{equation}
                                                        \label{Zp}
Z_p(T) = \mbox{Tr}_{S^1}\exp [-i{\cal H}_pt],
\end{equation}
can be expressed in terms of $\Gamma_p$ in the following
normalized form:
\begin{equation}
                                                   \label{ZpGamma}
Z_p(T) = \mbox{Tr}_{S^1}\Gamma_p(t,a)/\mbox{Tr}_{S^1} 1,
\end{equation}
where $\mbox{Tr}_{S^1}$ denotes the path integral over all the
$a$'s, which are $T$-periodic solutions to Hamilton's equation.

 The problem in computing $Z_p(T)$ given by (\ref{Zp})
arises due to the fact that the integral over all the $a$'s does
not converge.
 Finite value can be evaluated when realizing that the integral
receives contributions from the homotopy classes of the $a$'s
mentioned above so that we need to subtract excessive
degrees of freedom due to (\ref{ZpGamma}).

 Below, we perform explicit computation by realizing
$M^{2n}$ as a covering space, $f: M^{2n} \rightarrow Y^{2n}$, of
a suitable linearly connected manifold $Y^{2n}$ having the same
fundamental group as it of $S^1$,
$
\pi_1(Y^{2n}, t_0) \simeq \pi_1(S^1,t_0)
$.
 Since $\pi_0(Y^{2n}, t_0) = 0$ the set of preimages is
{\sl discrete},
\begin{equation}
                                                       \label{f-1}
f^{-1}(t_0) = \{ a_0, a_1, \dots \}.
\end{equation}
 The number of elements of $f^{-1}(t_{0})$ and of the monodromy
group given by the factorization,
\begin{equation}
                                                  \label{monodromy}
G =\pi_1(Y^{2n}, t_0)/f_{*}\pi_1(M^{2n}, a_0),
\end{equation}
coincides due to canonical one-to-one correspondence between
the set $f^{-1}(t_0)$ and the monodromy group $G$,
and does not depend on $t_0$ due to $\pi_0(Y^{2n}, t_0) = 0$.
Hence,
\begin{equation}
                                                        \label{ZpG}
Z_p(T) = \sum\limits_{g\in G}\Gamma_p(T, a_g),
\end{equation}
which is finite if $G$ is a finite group.

 Clearly, the $Z_p$'s are topological entities, which can be
used to define {\sl topological} entropy\cite{GR-92} of Hamiltonian
systems.
 Also, in terms of Morse theory it has been shown\cite{GR-90a,G-93}
that the partition function $Z(T)$ at $T\to 0$ localizes to critical
points of $H$ the number of which is given by (\ref{crit}) so that
the sum $\sum(-1)^pZ_p$ is the number of $T$-periodic solutions
to Hamilton's equation and is equal to Euler characteristic of
$M^{2n}$.
 Moreover, recent studies\cite{Niemi-94} showed that
if Hamiltonian $H$ is a {\sl perfect} Morse function on $M^{2n}$,
for which case one has to have $H^{2k+1}(M^{2n})=0$,
it saturates the lower bound in the Arnold conjecture, which
states that the number of nondegenerate contractible $T$-periodic
solutions of Hamilton's equation is greater or equal to sum of
Betti numbers of $M^{2n}$.

 In regard to the case of the symplectic tensor $\omega_{ij}$
depending on phase space coordinates considered in Sec.~\ref{Dfrm},
we have every reason to believe that the BRST observables and the
correlation functions will remain essentially of the same form
because they are intimately related to topology, which is
insensitive to the implemented differential (symplectic) structure.

\section{DISCUSSION}\label{discussion}

 After having analyzed the main ingredients of the construction
we make comments on the obtained results and discuss briefly on
the open problems and further developments.

 As we have already mentioned in Sec.~\ref{brst},
the symmetries of the resulting Hamilton function (\ref{H})
appeared to be even more than the BRST and anti-BRST symmetries we
have demanded upon.
 Indeed, the action $I$ is invariant under the ISp(2)
symmetry generated by the charges (\ref{QCK}).
 So, the reason of the occurrence of the additional symmetries,
$K$, $\bar K$, and $C$, should be clarified, in the context of
the BRST approach. For sure, these symmetries are
natural and establish the Poincare integral invariants,
as the fundamental topological observables, its conjugates,
and the ghost-number conservation.
Supersymmetry and also some of the above symmetries might be
broken by the term (\ref{Ipsi}), the role of which should be
investigated in a more detail.

 There are many directions worth pursuing. Probably the most
interesting are the following.

 (i) It is interesting to develop BRST approach to the theory
with explicit accounting for conservation of Hamiltonian,
$\dot H = 0$.
 Due to the fact that the $(2n-1)$-dimensional submanifold,
$M^{2n-1} \subset M^{2n}$, of constant energy, $H(a) = E$,
is invariant under the Hamiltonian flow, and the $p$-forms
evolve to $p$-forms on $M^{2n-1}$\cite{GR-92}, one can treat this
as a "reducible" action of the symplectic diffeomorphisms,
for which case more refined Batalin-Vilkovisky gauge fixing
formalism (see for a review Ref.\cite{Henneaux-85}) can be
applied, instead of the usual BRST one used in this paper.
 In general, the problem is to construct cohomological theory
for {\sl reduced} phase space of the model.

 The submanifolds $M^{2n-1}$, forming a one-parameter family,
have a rich set of possible topologies, depending on the
value of $E$, so that more refined analysis can be made on
generic Hamiltonian systems, for example, on bifurcations
(of invariant Liouville tori) in the system.
 Indeed, this approach having a great deal of cohomology
might yield information regarding topology and topology changes
of the submanifolds.
 Studies on classifying topologies of the constant energy
submanifolds are known in the mathematical literature.
 Particularly, in four-dimensional case ($n=2$), the submanifolds
for Hamiltonian systems having Bott integral of motion defined
on the submanifold has been studied by Fomenko
{\sl et al.}\cite{Fomenko-86}, who succeeded in complete
classification of possible topologies of the three dimensional
submanifolds for this case.

 Particular way to construct the effective theory is that one can
start with the theory with $M^{2n}$ as a target space.
 Next, implement the energy conservation constraint
$\delta (H(a)-E)$ to the path integral, that gives rise to
additional Lagrange multiplier, and introduce an auxiliary field
to constrain the flow to be tangent to the appropriately embedded
hypersurface $M^{2n-1}$, together with accompanying ghost and
anti-ghost fields.
 More geometrical way to account for the energy conservation
might be to extend the phase space by local coordinates
$x^{n+1} = t$ and $p^{n+1} = E$, and define the extended symplectic
two-form $\omega' = \omega -dt\wedge dE$.
 Then, the associated Hamilton's equation for the extended
Hamiltonian $H'= H-E$, on the hypersurface defined by the equation
$H'= 0$, is equivalent to the ordinary Hamilton's equation on
$M^{2n}$ plus two equations, $\dot t = -\partial H'/\partial E$
and $\dot E = \partial_t H$.
 Having such an extended formulation one might rerun
appropriately modified BRST procedure made in Sec.~\ref{brst}.

 One of the remarkable points we would like to note here is the
following. The constraint in the form $H(a)-\lambda E=0$ with
$\lambda$ viewed as a gauge parameter, can be accounted for to
obtain the theory with gauge fixed symmetry in respect to
transformation $\delta\lambda$.
 This corresponds exactly to the so called scaling systems,
{\sl e.g.} billiards, that have the same dynamics at all
energies and have received most of the attention so far
because they are easier to analyze and in many cases display
hard chaos.

(ii) More general approach to the above problem is to develop
a general functional scheme for {\sl Poisson manifold} instead of
symplectic one by relaxing the closeness and nondegeneracy
conditions for two-form $\omega$.
 The reason is that Poisson bracket, which is a central point
of consideration in this case, may be degenerate, for example,
for constrained systems, so one is led to study symplectic
shelves of Poisson manifold\cite{Karasev-91} on which Hamiltonian
dynamics is well defined and easier to treat.
 Classical Lie-Cartan reduction of the phase space and celebrated
theorem of non-commutative integrability (KAM theory) are specific
examples of such an approach.
Also, notice that Dirac bracket formalism is used to restore
Poisson brackets from known symplectic shelves defined by integrals
of motion.

 In general, this leads to consideration of nonlinear and/or
degenerate Poisson brackets, which are in fact most worth to study
since many systems reveals such a Poisson structure (after or even
without reduction of their phase space);
 for example, oscillator, pendulum, Euler rotations of rigid body,
spin dynamics of B-phase of superfluid $^3He$, and systems described
by classical Yang-Baxter equation.
 Also, we note that due to (iib) of Sec.~\ref{Dfrm}, in the
degenerate case, even small deformations of Poisson bracket
may cause global changes in topology of symplectic shelves
(bifurcations).

 The second reason of importance to develop functional approach to
dynamics on Poisson manifold is that just Poisson bracket is a
subject for usual and deformational quantization.

 (iii) We note that there is a tempting possibility to start with a
nontrivial topologically invariant underlying action $I_0$,
if it exist, instead of the trivial one.
 The problem is to construct an appropriate nontrivial topological
underlying Lagrangian ${\cal L}_0$, if any, for which the action
$I_0$ will not be dependent on the metric on $M^1$, a positive
definite function of time, {\sl einbein} $g = g(t)$, that is,
$\delta g = $ {\sl arbitrary}, $\delta I_0 = 0$.
 Such an invariance of the total action would be a kind of
time reparametrization invariance, $t\to\exp[\phi(t)]t$, and, in
fact, means coupling of the model to one-dimensional gravity
(see, for example, Ref.\cite{Marcus-94}).

 We note in this regard that one may be led to {\sl localize} the
BRST symmetry (\ref{s}).
 To do this, one can introduce gauge field $\eta$ with ghost number
-1, associated ghost $b$ with ghost number zero, and define
local version of the BRST operator, $s_l$, with the closed BRST
algebra being of the form
$s_la^i= bc^i$, $s_lc^i=0$, $s_l\bar c_i = bq_i$, $s_lq_i=0$,
$s_l\eta=-\partial_t b$, $s_l b=0$. Then, replacing $\partial_t$
in the gauge fermion (\ref{B}) by the BRST covariant operator,
$\partial_t-\eta s$, one arrives at the locally BRST invariant
Lagrangian, ${\cal L}=s_lB$. Simple calculations shows that the
ghost $b$ appears as an overall
factor and thus can be got rid of by rescaling $B$, with the
resulting Lagrangian being of the same form as (\ref{L}) minus
$iq_i\eta c^i$.
 The superspace interpretation of the gauge field $\eta$ is quite
clear, namely, it is a mixed component of the superspace metric,
$dx^2 = g^2dt^2 + \eta dt d\theta_1 + d\theta_1d\theta_2$.
 So, when requiring the metric independence of the total action,
one may insist on the independence on the gauge field $\eta$ as
well. It is highly remarkable to note that the latter may impose
nontrivial constraints on the form of Hamiltonian vector field
since it is the only auxiliary field in the theory outside of the
supermultiplet.

 (iv) Ultimately, of course, one would like go further in the
analysis of the $d=1$, $N=2$ supersymmetric model.
 One of the interesting problems, which escaped consideration in
this paper, and is presumably of much importance is geometry
of supersymmetric ground states, forming a space on the
coupling parameters entering (\ref{pert}).
 The metric of the Ramond ground states, $g_{i\bar j}$, is used to
extract interesting information on the physics, and satisfies the
topological-antitopological ($t\bar t$) equations\cite{Cecotti-91}.
In many cases they reduce to a familiar equation of mathematical
physics.
 It seems that valuable information can be obtained when analyzing
$t\bar t$ equations for the $d=1$, $N=2$ model under consideration,
for which we have shown that it admits Landau-Ginzburg
description.
 An example of the type of questions that we might want to
understand in the context of classical dynamical systems is, what
is the model where the same equations as the $t\bar t$ ones,
for this case, appear naturally. The difference from the
known analysis of the $t\bar t$ equations, both in $d=1$
and $d=2$ cases\cite{Cecotti-92}, may arise because not every
symplectic manifold admits a Kahler structure.

 (v) Further development can be made along the line of the phase
space formulation of ordinary quantum mechanics originated by
Weyl, Wigner and Moyal\cite{Weyl-Wigner-Moyal}.
 The key point one could exploit here is that it is
treated as a {\sl smooth
$\hbar$-deformation}\cite{GR-92d} of the classical mechanics
(see also \cite{JP-75}-\cite{Aringazin-94-JP}).
 Indeed, there is an attractive possibility to give an explicit
geometrical BRST formulation of the model describing quantum
mechanics in phase space, following the lines of the present paper.
 The resulting theory could be thought of as a topological phase
of quantum mechanics in phase space.
 The crucial part of the work has been done\cite{GR-92d} in the
path integral formulation, where the associated extended phase space
and quantum $\hbar$-deformed exterior differential calculus in
quantum mechanics has been proposed.
 The core of this formulation is in the deforming of the Poisson
bracket algebra of classical observables.

 The central point we would like to use here is that the extended
phase space can be naturally treated as the cotangent superbundle
$M^{4n|4n}$ over $M^{2n}$ endowed with the second symplectic
structure $\Omega$ and graded Poisson brackets (\ref{epb}).
 Besides clarifying the meaning of the ISp(2) algebra, appeared as a
symmetry of the field theoretic model, it allows one, particularly,
to combine symplectic geometry and techniques of fiber bundles.
 The underlying reason of our interest in elaborating the fiber
bundle construction is that one can settle down Moyal's
$\hbar$-deformation in a {\sl consistent} way by using both of the
Poisson brackets, $\{,\}_\omega$ and $\{,\}_\Omega$. Namely, the
two symplectic structures and Hamiltonian vector fields coexisting
in the single fiber bundle are related to each
other\cite{Aringazin-95}.
 Note that this relation is not direct since
$\{a^i,a^j\}_\omega=\omega^{ij}$ while for the projection of
coordinates in the fundamental Poisson bracket
$\{\lambda^a,\lambda^b\}_\Omega=\Omega^{ab}$ to the base $M^{2n}$
we have $\{a^i,a^j\}_\Omega=0$.
 Also, $Z_2$ symmetry (\ref{Z2}) of the undeformed Lagrangian
(\ref{L'}) can be used as a further important requirement for the
deformed extension.
 Naively, the problem is to construct $\hbar$-deformed BRST exact
Lagrangian, identify BRST invariant observables, and study BRST
cohomology equation and corresponding correlation functions.
 Also, having the conclusion that the $d=1$, $N=2$ supersymmetry
plays so remarkable role in the classical case it would be
interesting to investigate its role in the quantum mechanical case.

 In this way, one might formulate, particularly, quantum analogues
of the Lyapunov exponents (\ref{Lyapunov}) in terms of correlation
functions rather than to invoke to nearby trajectories, which make
no sense in quantum mechanical case. The case of compact classical
phase space corresponds to a finite number of quantum states.
 Also, we note that for chaotic systems expansion on the periodic
orbits constitutes the only semiclassical quantization scheme known.
Perhaps, this is a most interesting problem, in view of the recent
studies of quantum chaos.

 However, we should emphasize here that the geometrical BRST
analogy with the classical case is not straightforward, as it may
seem at first glance, since one deals with non-commutative
geometry\cite{Connes-93} of the phase space in quantum mechanical
case (see Ref.\cite{Aringazin-94-JP} and references therein).
 Particularly, quantum mechanical observables of interest are
supposed to be analogues of the closed $p$-forms on $M^{2n}$,
with noncommuting coefficients arising to nonabelian cohomology.

\section*{ACKNOWLEDGEMENTS}

 The authors are much grateful to E.Gozzi for many valuable
comments on preliminary version of the manuscript.
 A.K.A. is grateful to M.Majitov and A.Kurilin for helpful
discussions.

 This work is supported in part by Science Foundation of
the Ministry of Science of Kazakstan, grant N104-96.

\appendix
\section{APPENDIX}

 In Appendix (see also ref.\cite{Aringazin-94-HJ}),
we obtain explicitly solutions of the
supersymmetric ground state equations, both in Hamiltonian and
Birkhoffian cases, for two-dimensional phase space, $n = 1$, to illustrate
emerging
of the Gibbs distribution.

 In the phase space with Darboux coordinates, $a^1 = p$ and $a^2 = x$,
the non-vanishing coefficients of the symplectic tensor are given by
$\omega^{12} = -\omega^{21} = 1$.
 The general expansion of the ghost dependent distribution reads
\begin{equation}
 \rho (a,c) = \rho_{0}(a)+\rho_{1}(a)c^{1}+ \rho_{2}(a)c^{2}
                      +\rho_{12}(a)c^{1}c^{2}.
\end{equation}
 In general, each ghost sector in $\rho(a,c)$ can be used to define
some ordinary type of distribution.
 The ground state equations (\ref{Qbetarho}) then read
\begin{eqnarray}
                                                         \label{r0}
c^1(\partial_1 - \beta h_1)\rho_0 = 0, \quad
c^2(\partial_2 - \beta h_2)\rho_0 = 0,\\
                                                         \label{r12}
c^1(\partial_1 + \beta h_1)\rho_{12} = 0, \quad
c^2(\partial_2 + \beta h_2)\rho_{12} = 0,\\
                                                         \label{r1r2}
c^1c^2\bigl[(\partial_1 - \beta h_1)\rho_2
          - (\partial_2 - \beta h_2)\rho_1)\bigr] = 0,\\
                                                         \label{r2r1}
(\partial_1 + \beta h_1)\rho_2 -
(\partial_2 + \beta h_2)\rho_1 = 0.
\end{eqnarray}
 Here, $\partial_i \equiv \partial /\partial a^i$ and
$h_i \equiv \partial H(a^1,a^2) /\partial a^i$ $(i = 1, 2)$.
 For the ghost-free sector (\ref{r0}) and the two-ghost sector
(\ref{r12}) we have immediately
\begin{eqnarray}
\rho_0= A_0(a^2)\exp[\beta H], \quad
\rho_0= B_0(a^1)\exp[\beta H],\\
\rho_{12} = A_{12}(a^2)\exp[-\beta H], \quad
\rho_{12} = B_{12}(a^1)\exp[-\beta H]
\end{eqnarray}
where $A$'s and $B$'s are arbitrary functions, with the general solutions
\begin{eqnarray}
\rho_0= const\exp[\beta H],\\
\rho_{12} = const\exp[-\beta H].
\end{eqnarray}
 These ghost sectors define scalar and pseudoscalar distributions,
$\rho_s=\rho_0$ and $\rho_{ps}=\rho_{12}da^1\wedge da^2$,
respectively.

 The equations (\ref{r1r2}) and (\ref{r2r1}) can be rewritten as
\begin{equation}
\partial_1 \rho_2 - \partial_2 \rho_1 = 0, \quad
2\beta (h_1 \rho_2 - h_2 \rho_1) = 0,
\end{equation}
or, taking $\beta > 0$,
\begin{equation}
                                                       \label{r1}
\frac{h_2}{h_1}\partial_1 \rho_1 - \partial_2 \rho_1
        = - \rho_1\partial_1\frac{h_2}{h_1}, \quad
\rho_2  = \frac{h_2}{h_1}\rho_1.
\end{equation}
 To solve the nonhomogeneous first-order partial differential
equation (\ref{r1}) for $\rho_1$, we write down, by a standard
technique, its characteristic equations,
\begin{equation}
\frac{da^1}{dr} = \frac{h_2}{h_1}, \quad
\frac{da^2}{dr} = - 1,             \quad
\frac{d\rho_1}{dr} = - \rho_1\partial_1\frac{h_2}{h_1},
\end{equation}
where $r$ is a parameter, from which the first and the second
integrals follow,
\begin{equation}
 U_1 = \int (h_1 da^1 + h_2 da^2), \quad
 U_2 = \frac{h_2}{h_1}\rho_1.
\end{equation}
The general solution is then of the form $\Phi(U_1, U_2) = 0$,
where $\Phi$ is a function, that is, we can write
\begin{equation}
                                                     \label{r1sol}
\rho_1 = \frac{h_1}{h_2}f(U_1),
\end{equation}
and hence $\rho_2 = f(U_1)$, where $f$ is an arbitrary function.
Symmetrically, one can arrive at the solutions in the form
$\rho_1=f(U_1)$ and $\rho_2=(h_2/h_1)f(U_1)$.
 Geometrically, the odd-ghost sectors $\rho_1$ and $\rho_2$
constitute the vector distribution, $\rho_v=\vec\rho(a)d\vec a$.

 To clarify the possible meaning of such a distribution we make
some comments.
 It is clear that this distribution is not of a Gibbs form, does not
depend on the parameter $\beta>0$, and singular at the
critical points of the gradient vector field $\vec h\equiv(h_1,h_2)$.
 The latter implies that $\rho_v$ is not in general normalizable.
 However, in the region that does not include critical points of
$\vec h$ the distribution $\rho_v$ is well defined.
 When, additionally, we specify the integrating in $U_1$ to be
over a closed path $\partial D$ we have $U_1=0$ identically since
$U_1=\int_{\partial D}\vec h\, d\vec a
=\int_D\mbox{rot}\vec h\, d\vec\sigma
=\int_D\mbox{rot}\,\mbox{grad}H\, d\vec\sigma \equiv 0$.
Hence, the solution (\ref{r1sol}) reduces to
$\rho_1=h_1/h_2$ and $\rho_2=$ const.

In the Birkhoffian case, $\omega^{12}(a) = -\omega^{21} = \omega(a)$,
we have, instead of equations (\ref{r12}), the following:
\begin{equation}
\label{Br12}
c^1(\omega(\partial_1 + \beta h_1)\rho_{12} + \partial_1\omega)= 0, \quad
c^2(\omega(\partial_2 + \beta h_2)\rho_{12} + \partial_2\omega)= 0,
\end{equation}
while the other equations are of the same form (\ref{r0}),
(\ref{r1r2}), and (\ref{r2r1}). It is remarkable that dependence
of $\omega$ on $a^i$ display itself only for $\rho_{12}$, in which we are
most interested. Solutions of the above equations are:
\begin{equation}
\label{Sol1Br12}
\rho_{12}= \exp[-\beta H]\bigl[ B_{12}(a^2) -
\int \frac{\exp[\beta H]\partial_1\omega}{\omega}da^1\bigr],
\end{equation}
\begin{equation}
\label{Sol2Br12}
\rho_{12} =\exp[-\beta H]\bigl[ A_{12}(a^1) -
\int \frac{\exp[\beta H]\partial_2\omega}{\omega}da^2\bigr].
\end{equation}

\end{document}